\documentclass[11pt,a4paper]{article}

\usepackage[margin=1in]{geometry}
\usepackage{graphicx}%
\usepackage{amsmath,amssymb,amsfonts}%
\usepackage{xcolor}%
\usepackage{booktabs}%
\usepackage{caption}
\usepackage{subcaption}
\usepackage{makecell}
\usepackage{siunitx}
\usepackage{authblk}

\usepackage{algorithm}%
\usepackage{algorithmicx}%
\usepackage{algpseudocode}%

\usepackage[numbers,sort&compress]{natbib}
\makeatletter
\newcommand{\equalcontrib}{\textsuperscript{*}}
\newcommand{\corrauth}{\textsuperscript{\dag}}
\newcommand{\printauthornotes}{%
  \begingroup
  \renewcommand\thefootnote{\fnsymbol{footnote}}
  \footnotetext[1]{These authors contributed equally to this work.}%
  \footnotetext[2]{Corresponding author: \texttt{christian.salomonsen@uit.no}}%
  \endgroup
}
\makeatother

\newcommand{\fdg}{\ensuremath{\mathrm{[^{18}F]FDG}}}
\newcommand{\fdopa}{\ensuremath{\mathrm{[^{18}F]FDOPA}}}
\newcommand{\psma}{\ensuremath{\mathrm{[^{68}Ga]PSMA}}}

\DeclareSIUnit{\minutes}{minutes}
\DeclareSIUnit{\weeks}{weeks}
\DeclareSIUnit{\hours}{hours}

\title{A robust and versatile deep learning model for prediction of the arterial input function in dynamic small animal $\left[^{18}\text{F}\right]$FDG PET imaging}

\author[1]{Christian Salomonsen\equalcontrib\corrauth}
\author[1]{Luigi T. Luppino\equalcontrib}
\author[1]{Fredrik Aspheim}
\author[1]{Kristoffer K. Wickstr{\o}m}
\author[1]{Elisabeth Wetzer}
\author[1]{Michael C. Kampffmeyer}
\author[2,3]{Rodrigo Berzaghi}
\author[2,3]{Rune Sundset}
\author[1]{Robert Jenssen}
\author[1,2]{Samuel Kuttner}

\affil[1]{Department of Physics and Technology, UiT The Arctic University of Norway, Troms{\o}, Norway}
\affil[2]{PET Imaging Center, University Hospital of North Norway, Troms{\o}, Norway}
\affil[3]{Department of Clinical Medicine, UiT The Arctic University of Norway, Troms{\o}, Norway}

\date{\today}

\begin{document}

\maketitle
\printauthornotes

\begin{abstract}
  Dynamic positron emission tomography (PET) and
  kinetic modeling are pivotal in advancing tracer development research in small
  animal studies. Accurate kinetic modeling requires precise input function
  estimation, traditionally achieved through arterial blood sampling. However,
  arterial cannulation in small animals, such as mice, involves intricate,
  time-consuming, and terminal procedures, precluding longitudinal studies. This
  work proposes a non-invasive, fully convolutional deep learning-based approach
  (FC-DLIF) to predict input functions directly from PET imaging data, which may
  eliminate the need for arterial blood sampling in the context of dynamic
  small-animal PET imaging.

  The proposed FC-DLIF model consists of a spatial feature
  extractor that acts on the volumetric time frames of the dynamic PET imaging
  sequence, extracting spatial features. These  are subsequently further processed
  in a temporal feature extractor that predicts the arterial input function. The
  proposed approach is trained and evaluated using images and arterial blood
  curves from \fdg{} data using cross validation. Further, the model applicability
  is evaluated on imaging data and arterial blood curves collected using two
  additional radiotracers (\fdopa{}, and \psma{}). The model was further evaluated
  on data truncated and shifted in time, to simulate shorter, and shifted, PET
  scans.

  The proposed FC-DLIF model reliably predicts the arterial
  input function with respect to mean squared error and correlation. Furthermore,
  the FC-DLIF model is able to predict the arterial input function even from
  truncated and shifted samples. The model fails to predict the AIF from samples
  collected using different radiotracers, as these are not represented in the
  training data. 

  Our deep learning-based input function offers a
  non-invasive and reliable alternative to arterial blood sampling, proving robust
  and flexible to temporal shifts and different scan durations. 
\end{abstract}

\clearpage

\section{Introduction}
\label{src:introduction}

Dynamic positron emission tomography (PET) plays a critical role in the imaging of small
animals, enabling \textit{in vivo} visualization of tracer uptake over time, and
subsequent quantification of tracer transport or binding through kinetic
modeling~\cite{Gunn2001b}. This is actively used as a tool for downstream tasks, for
example in the development of new tracers, drugs, diagnostic procedures, and disease
therapies~\cite{hicks2006pet, yao2012small, cunha2014preclinical}. However, a
prerequisite for kinetic modeling is knowledge of the arterial tracer concentration over
time, commonly referred to as the arterial input function (AIF)~\cite{Alf2013e}. 

Arterial blood sampling is considered the gold standard for AIF acquisition, but in
preclinical settings it presents serious limitations. The procedure is technically
complex, time-consuming, and terminal in mice due to the invasiveness of carotid
cannulation. Furthermore, only a limited amount of blood volume can be withdrawn without
altering animal physiology~\cite{Laforest2005,Convert2022}. These limitations make
high-throughput and longitudinal studies infeasible and may further raise ethical
concerns regarding excessive animal usage~\cite{russell1959principles}.

This has driven significant interest in non-invasive AIF estimation strategies.
Population based input functions (PBIFs) averages time-activity curves (TACs) from
demographically similar subjects~\cite{takikawa1993PBIF}, but fail to capture individual
variability and still requires at least one blood sample to scale the population curve.
Image-derived input functions (IDIFs) extract TACs from vascular regions such as the
left ventricle, but are prone to inaccuracies stemming from partial-volume effects,
motion artifacts, and poor signal-to-noise
ratios~\cite{zanotti2011IDIFchallenges,Laforest2005,frouin2002correction,kim2013partial,fang2008spillover}.
Simultaneous estimation (SIME) approaches reduce the reliance on blood sampling by
fitting both tissue kinetics and the AIF from multiple regions
concurrently~\cite{van2023noninvasive}, but they require a predefined functional form
for the AIF and still depend on at least one late-time blood sample to anchor the
estimated curve to absolute tracer
concentrations~\cite{bartlett2019quantification,roccia2019quantifying,feng1997SIME,wong2001SIME}.

To address these limitations, several data-driven alternatives have been proposed. In
our previous research, we developed machine learning-based methods for AIF estimation
using extracted TACs as input~\cite{Kuttner2020,Kuttner2021}. While these methods
eliminated the need for blood sampling, they relied on manually delineated regions of
interest (ROIs) and post-imaging TAC extraction, limiting their scalability and
integration into automated pipelines.

In more recent studies, it has been demonstrated that deep learning-based methods
outperform more traditional machine learning-based methods bypassing the need for
handcrafted features and manual ROI delineation by prediction of the AIF directly from
the dynamic PET image
volume~\cite{wang2024noninvasive,kuttner2024deep,ferrante2024physically,varnyu2021DLIF}.
However, most of these methods adopt hybrid architectures that combine 3D convolutional
layers with fully connected or recurrent layers. These architectural choices introduce
key limitations. For example, fully connected layers necessitate fixed-length inputs,
forcing all dynamic scans to be padded, truncated, or interpolated to a uniform number
of time frames prior to inference~\cite{wang2024noninvasive,kuttner2024deep}. This
increases computational overhead and may introduce interpolation artifacts, especially
when acquisition protocols differ in temporal resolution or duration.

Recurrent-style models, such as LSTM-based
networks~\cite{wang2024noninvasive,varnyu2021DLIF}, attempt to model temporal
dependencies explicitly, but often learn features tied to specific time indices,
reducing robustness to timing shifts or variation in tracer arrival across subjects.
Furthermore, several methods include \textit{post hoc} fitting or model parameter
regression~\cite{varnyu2021DLIF,ferrante2024physically}, adding assumptions and
complexity to the pipeline. These constraints can hinder generalization and reduce
flexibility in practical settings.

In this study, we propose a novel approach---fully convolutional deep learning-based
input function prediction (FC-DLIF)---that overcomes these limitations by using only
convolutional operations over both spatial and temporal dimensions. The fully
convolutional design allows FC-DLIF to process input sequences of arbitrary length,
eliminating the need for dense or recurrent layers, and removing dependence on fixed
input shapes. FC-DLIF takes reconstructed 4D dynamic PET data (t, x, y, z) and directly
predicts an AIF output, with no need for manual ROI segmentation, TAC extraction,
temporal resampling, or post hoc fitting. By treating time as a learnable axis and
detecting temporal patterns such as peak onset, plateau, and washout tails, regardless
of absolute frame positions, FC-DLIF is inherently robust to timing variability and
flexible across different imaging protocols. The architecture provides a streamlined,
end-to-end pipeline for non-invasive AIF estimation suitable for diverse preclinical
study designs.

We evaluate FC-DLIF on a dynamic PET dataset of mice with paired arterial blood data,
and show that our method generalizes across protocols with time-shifted inputs and
different scan durations. By accurate, non-terminal, and non-invasive AIF estimation,
FC-DLIF has the potential to reduce animal use, facilitating longitudinal studies, and
simplify the workflow for kinetic-modeling, aligning with the 3Rs principles of animal
research~\cite{russell1959principles}.

\section{Materials and methods}

\begin{figure}[t]
    \centering
    \includegraphics[width=\textwidth]{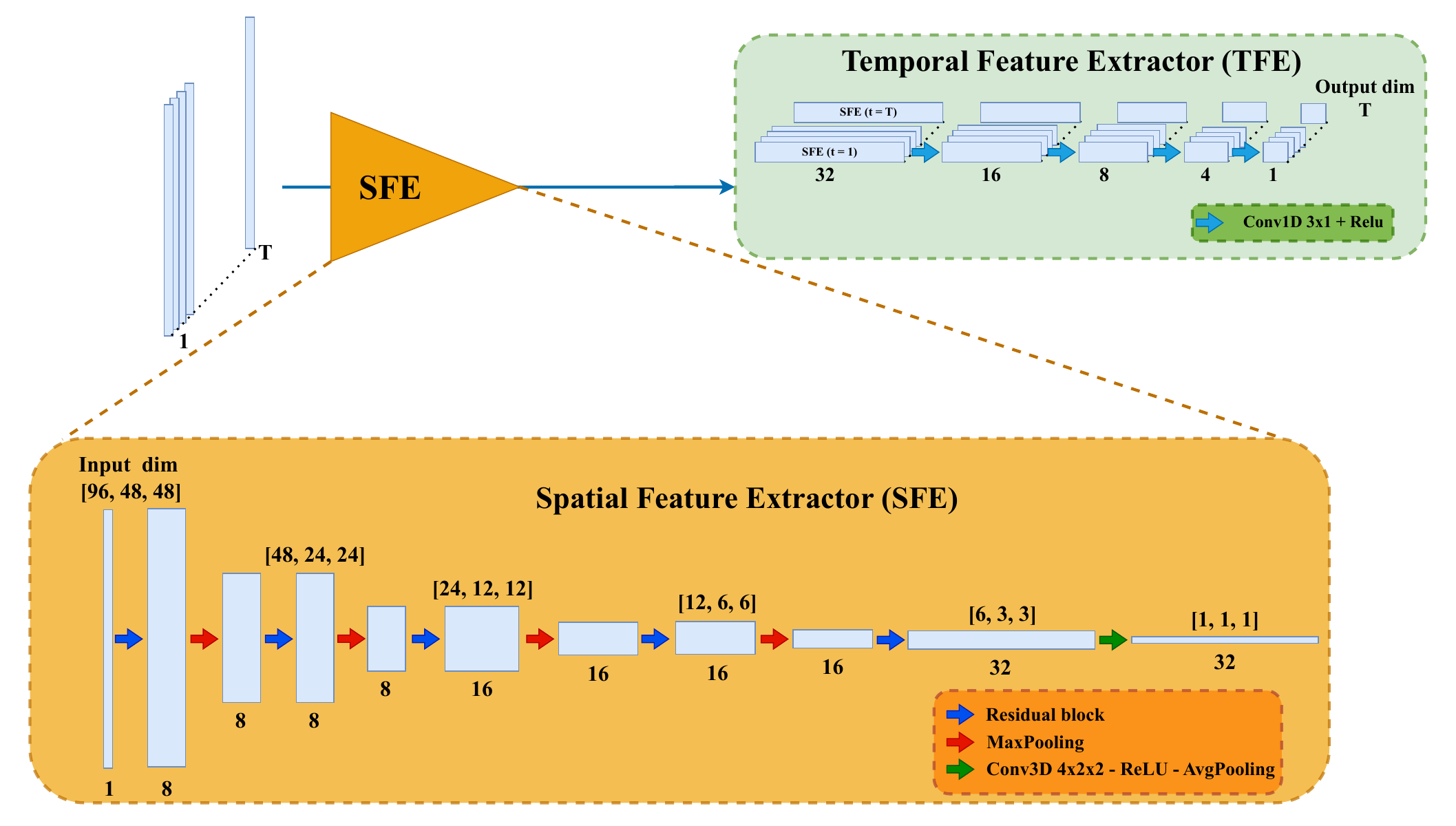}

    \caption[The proposed FC-DLIF architecture.]%
    {The proposed architecture consists of
    two parts: a 3D ResNet acts as a spatial feature extractor (SFE), and a 1D
    convolutional network acts as a temporal feature extractor (TFE). The first takes
    in input the data volume of each time step and reduces its dimensions to one
    vector of \num{32} features. Then the extracted vectors of all time steps are
    stacked along the time dimension, over which the second part of the model performs
    a series of convolutions, until the vector dimension is reduced to 1. The final
    output of the model is a time-series with the same length as the input data.}
    \label{fig:model}
\end{figure}

\label{sec:methods}

This section describes the data, the design of the proposed deep learning model, and its
training procedure. Further details on data collection are provided in
Appendix~\ref{sec:appendix-data}.

\subsection{Dataset}
\label{sec:data}

The training dataset consisted of \num{70} dynamic \fdg{} PET scans with
simultaneously measured AIFs of mice in ages \numrange{9}{24} weeks, collected
at UiT The Arctic University of Norway (UiT). The dataset included three
different mouse strains: BALB/cJRj (N = \num{55}), C57BL/6JRj (N = \num{8}) and
Balb/cAnNCrl (N = \num{7}). Another \num{10} samples, collected with different
tracers \fdopa{} (N = \num{6}) and \psma{} (N = \num{4}), which were not
included during model training, but used to evaluate the model performance on
new, unseen tracers. Following PET imaging, dynamic image volumes were
reconstructed into \num{42} time-frames
(\qtylist[parse-numbers=false]{\numproduct{1x30};\numproduct{24x5};\numproduct{9x20};\numproduct{8x300}}{\second})
of whole-body data with shape \numproduct{42x96x48x48}, and normalized into
standardized uptake value (SUV)~\cite{Keyes1995a}. 

\subsection{Model architecture}
\label{sec:model}

Fig.~\ref{fig:model} shows the architecture of the proposed model. It consists
of two parts, one spatial, and one temporal feature extractor, that serve
different purposes. First, the data volume associated with each time step of the
dynamic scans is passed through a 3D ResNet~\cite{resnet}, called a spatial
feature extractor (SFE), which extracts relevant spatial features while reducing
only the spatial dimensions. From the input, each data volume goes through a
series of residual blocks interleaved with max-pooling layers, until a
convolutional layer with a \numproduct{4x2x2} cuboid kernel, followed by an
adaptive average-pooling layer reduces said volume to a one-dimensional vector
of \num{32} features. This acts as a compact representation for each time step.
Effectively, each time frame is processed independently by the SFE, exposing the
same convolutional filters to each part of the dynamic image. In the second part
of the model, all the extracted vectors are stacked along the time dimension,
over which the temporal feature extractor (TFE) captures temporal correlations
between the time frames. The TFE uses 1D convolutions, which is motivated by
recent research that has demonstrated superior performance and efficiency
compared to alternative methods based on recurrent neural networks
\cite{timeseriesCNN}. The output of the TFE is the final output of the model and
represents the DLIF prediction of the AIF, given the data volume.

\subsection{Training procedure}

The ADAM optimizer~\cite{ADAM, reddi2018on} with standard settings was used to
minimize the weighted mean squared error (wMSE) between the predicted DLIF and
the ground truth. The error on each time step was weighted to account for the
imbalance between three parts of the curve: the first \num{25} (peak), middle
\num{9} (intermediate), and last \num{8} (tail) frames were associated with
weights \numlist{0.4;0.7;1} respectively. Training was performed with a learning
rate of \num{1e-4} for \num{1000} epochs. \num{10}-fold cross validation allowed
the evaluation of the model performance over the whole training dataset. For
each fold, \num{10} runs were repeated for statistical rigor. Data augmentation
with additive random Poisson noise injection were used during training, with the
goal of exposing the model to different signal-to-noise ratios and accustom it
to lower image qualities. The degraded image was generated by sampling a scalar
$p\sim \mathrm{Unif}(0,1)$, then, for a dynamic PET image with voxel intensities
$I$, the corresponding image with noise, $I_{\mathrm{poisson}}$ is defined as:

\begin{equation}
    \begin{aligned}
        \Lambda &= I \cdot p\\
        I_\mathrm{poisson} &\equiv I + \mathrm{Pois}(\Lambda) - \Lambda
    \end{aligned}
\end{equation}

\begin{figure}[t]
    \centering
    \begin{subfigure}[t]{0.48\columnwidth}
        \includegraphics[width=\linewidth,keepaspectratio]{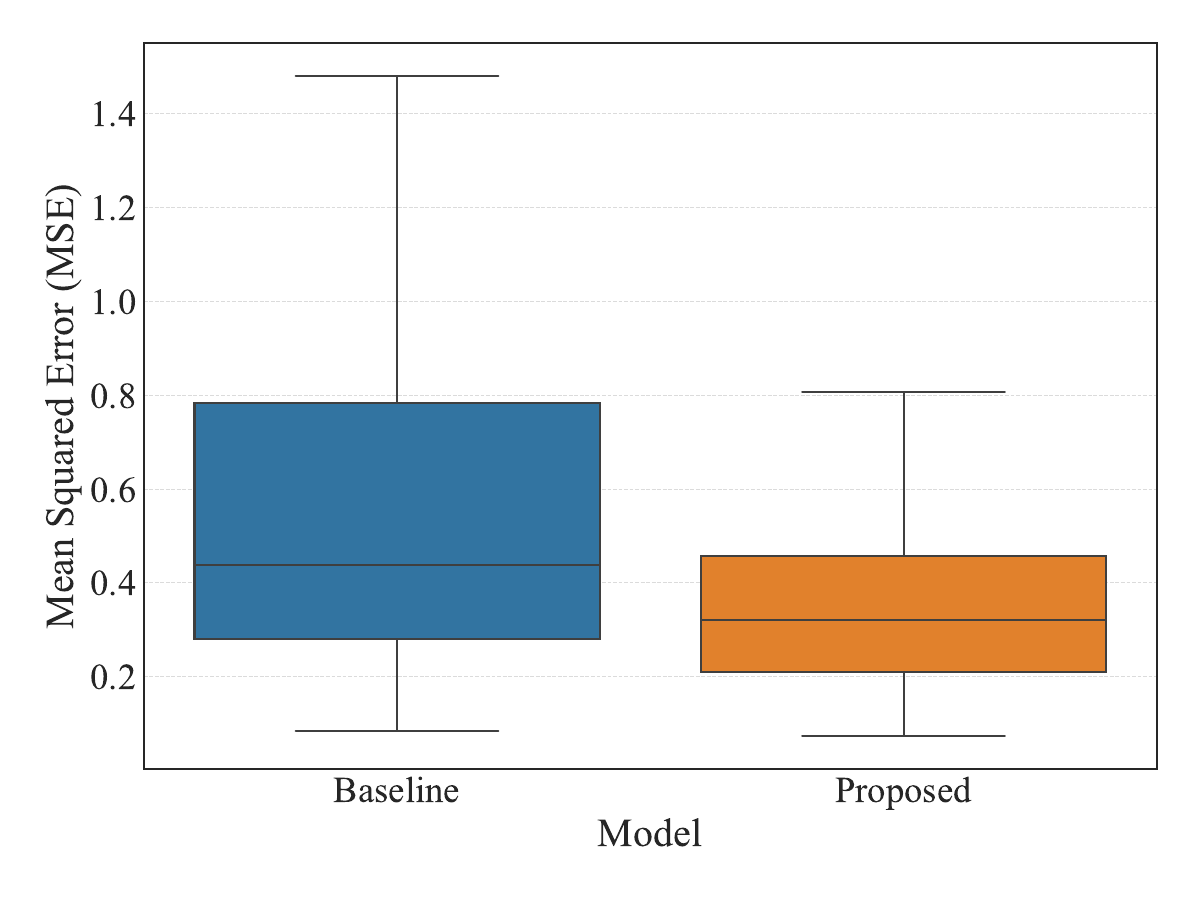}
        \caption{MSE boxplot}
        \label{fig:mse-boxplot}
    \end{subfigure}
    \begin{subfigure}[t]{0.48\columnwidth}
        \includegraphics[width=\linewidth,keepaspectratio]{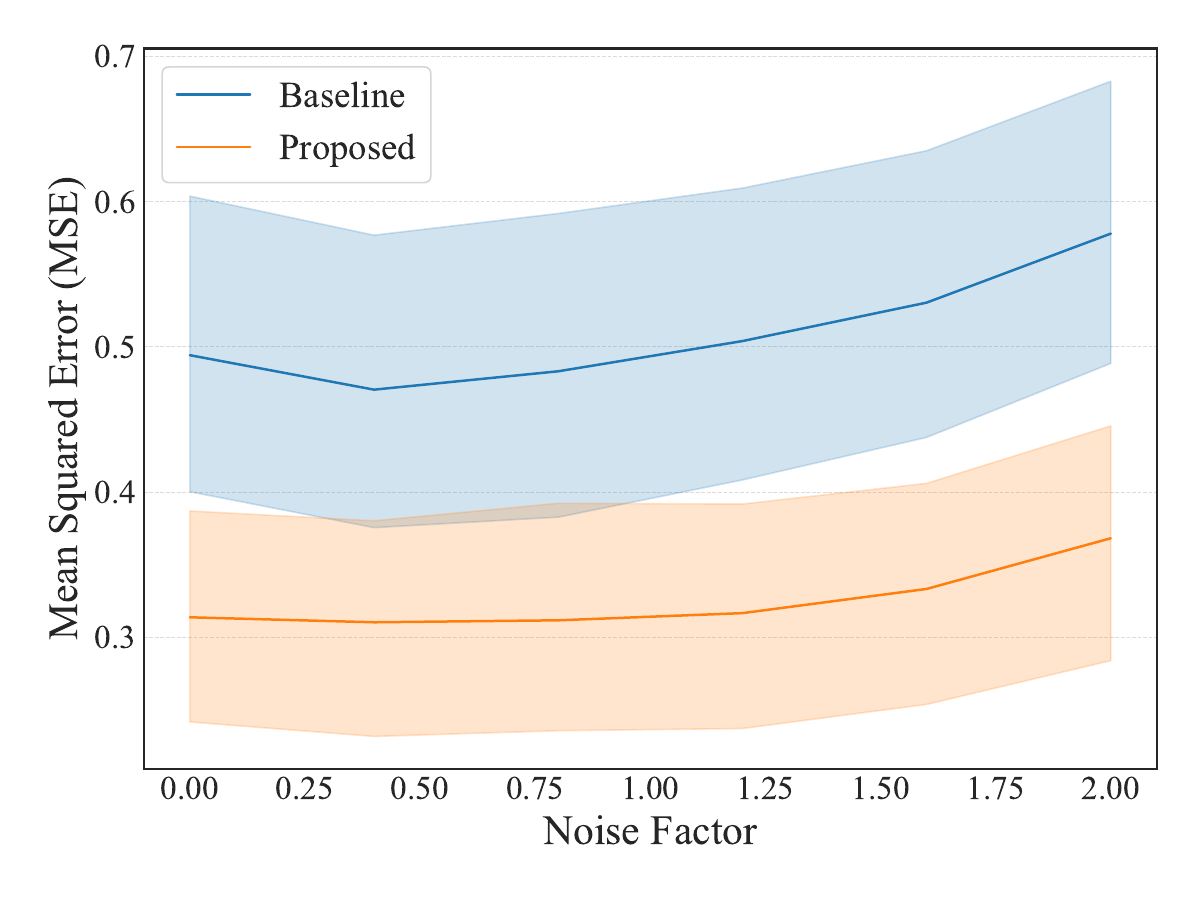}
        \caption{MSE over different injected noise factors}
        \label{fig:noise}
    \end{subfigure}

    \caption[Quantitative comparison between baseline and proposed model.]%
    {Quantitative comparison between the baseline and the proposed model over each
    sample, (a) boxplot of the MSE from the two models; (b) Lineplot with mean and 95\%
    confidence interval for different Poisson noise factors.}
    \label{fig:quantitative}
\end{figure}

\subsection{Evaluation metrics}

The predicted DLIF curves were compared time frame by time frame with the respective
measured AIF using a paired t-test ($\alpha = 0.05$), and orthogonal regression to
account for measurement errors in both predicted and measured variables. Normality was
assessed using quantile-quantile plots, reporting the Pearson correlation to describe
the spread of the predicted curves. In order to assess the efficacy of the proposed
method for downstream tasks, such as kinetic modeling, the graphical Patlak
plot~\cite{graphical1983patlak,graphical1985patlak} is used to produce net influx rates
K$_\mathrm{i}$ given the reference and predicted input functions for each voxel over
time. Prior to kinetic modeling, both the AIF and DLIF curves were converted to plasma
input functions following~\cite{wu2007plasmacalibration}. DLIF-based influx estimates
were further compared to those obtained from the measured AIF using the same evaluation
methods as for the predicted curves.

Comparisons were done against the DLIF model proposed by Kuttner et
al.~\cite{kuttner2024deep}, which will be referred to as \textit{the baseline}
in the following sections. To ensure fair comparisons, the baseline was trained
on the same data as the FC-DLIF, also using a \num{10}-fold cross validation
setting.

Additionally and unrelated to the above comparisons, the FC-DLIF network's
ability to operate on data with varying temporal dimensions were explored using
two tests. First, a time-shifted sample was simulated by prepending the initial
time frame to the dynamic PET image---before the radiotracer uptake phase had
begun---to reveal any learned time-specific dependencies. For instance, the
network may insist that time frame $I_t$ always contains the peak of the input
function during inference.

Second, the dynamic image was truncated by removing the first \num{4} frames
(first \qty{40}{\second}), and the last \num{6} time frames (last
\qty{30}{\minute}) of the signal. This simulates a change in the imaging
protocol, where radiotracer infusion and imaging is simultaneously started,
partially masking the input function onset, and ending the scan earlier. This
experiment was hypothesized to reveal the models ability to work with limited
scan durations, while retaining performance on the task.

To investigate the latent representations extracted from the SFE, the
t-SNE~\cite{van2008visualizing} algorithm was used. The algorithm was applied to
one model from a single run of a specific fold (fold 2, run 9) to visualize
common traits of the condensed spatial representation.

\section{Results}

\begin{figure}[t]
    \centering
    \includegraphics[width=\linewidth,keepaspectratio]{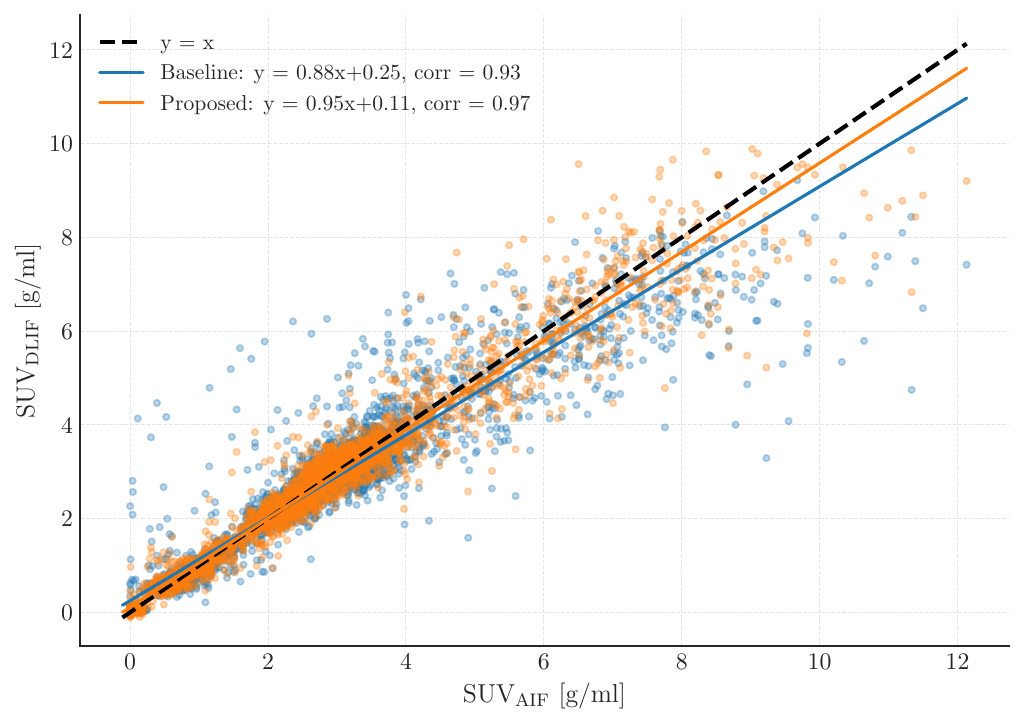}

    \caption[Scatterplot of AIF estimates.]%
    {Scatterplot summarizing the results over the whole dataset for the arterial input
    function estimation for each model. The points show the pointwise comparison
    between true and predicted values for each mouse, represented by \num{42} time
    steps for both the baseline and proposed models. The axes labels indicate the SUV
    for either the predicted (SUV$_\mathrm{DLIF}$), or measured input function
    (SUV$_\mathrm{AIF}$). Ideally for a perfect fit, all points would lie on the black
    dashed line $y=x$, with $a=1$ and $b=0$. The coefficient of determination $r^2\leq1$
    quantifies how well the predictor fits the data. Similarly, Pearson's correlation
    coefficient $r\leq 1$ measures the linear correlation between the AIF and the DLIF.}
    \label{fig:scatter_AIFs}
\end{figure}

First, this section summarizes the results of the proposed FC-DLIF model during
cross-validation on our dataset when compared against the baseline, both in terms of
errors on AIF estimation and, consequently, after tracer kinetic modeling. Then, some
examples of AIF estimation show the performance and versatility of the proposed model
with respect to the input data. Finally, the features extracted by the model's SFE is
visualized and inspected using the t-SNE algorithm~\cite{van2008visualizing}.

\subsection{Quantitative results}

Fig.~\ref{fig:quantitative} summarizes the distribution of mean squared error (MSE,
Fig.~\ref{fig:mse-boxplot}) and injected noise (Fig.~\ref{fig:noise}) across all samples.
FC-DLIF consistently achieved lower MSE values than the baseline, with a noticeably
narrower spread and a more favorable distribution across percentiles. In terms of noise,
FC-DLIF also demonstrated reduced variability, being less sensitive to noise in the
input data.

In particular, the proposed model has close to \SI{75}{\percent} of MSE vales
lower than the median of the baseline model, and generally produce predictions
more aligned with the AIF, as seen in the $5^{th}$ percentile (best
\SI{95}{\percent} of predictions), which is slightly lower than the baseline
(Fig.~\ref{fig:mse-boxplot}). These results are further consolidated when
comparing the same metrics over temporal regions, as shown in
Fig.~\ref{fig:curve phases}, where the proposed method exhibits less bias in
most segments.

In Fig.~\ref{fig:scatter_AIFs}, the predicted AIF values from each model are
compared against ground truth measurements on a point-wise basis. The proposed
method achieves higher $R^2$ and Pearson correlation values, with predictions
more closely aligned to the ideal $y = x$ line, indicating a tighter spread of
predictions around the regression line. While the baseline exhibits systematic
deviations, particularly overestimating intermediate and tail values and
underestimating peak amplitudes, FC-DLIF maintains a tighter spread around the
regression line.

Predictions from the baseline model are generally overestimating small values
(less than \SI{4}{g/ml}), and underestimating larger values, associated with the
peak of the input function curve. The same characteristics are not seen for the
FC-DLIF method, but instead the spread is generally larger above values of
\SI{4}{g/ml}, which is also shown in Fig.~\ref{fig:curve phases}, for the peak
labeled segments.

\begin{figure}[t]
    \centering
    \includegraphics[width=\linewidth,keepaspectratio]{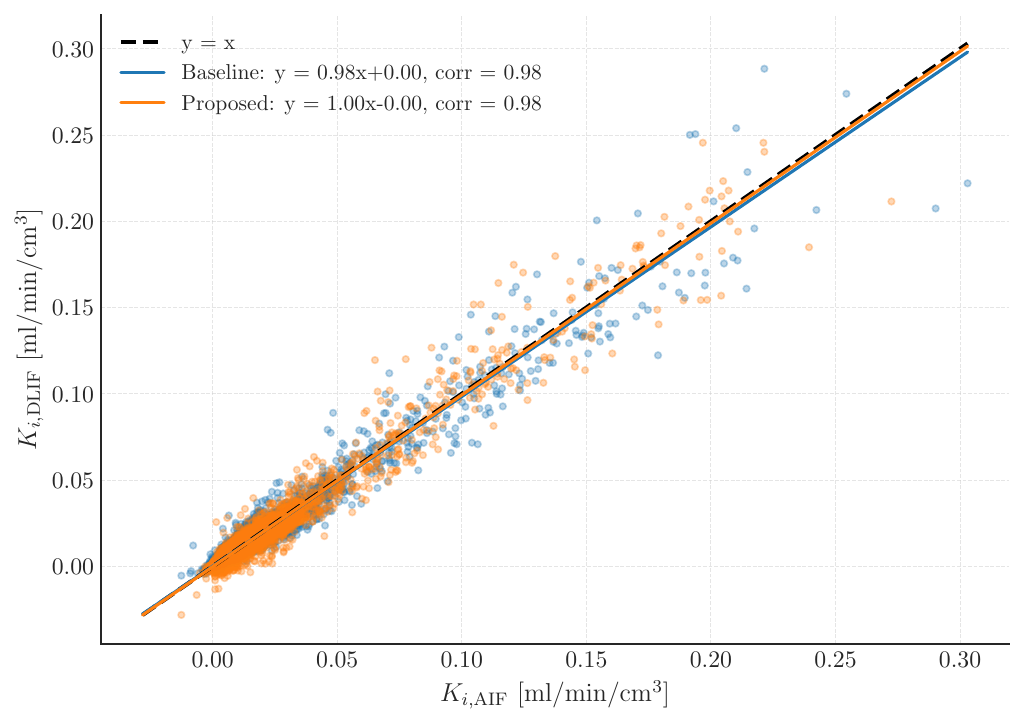}

    \caption[Scatterplot of kinetic parameters.]%
    {Scatterplot summarising the results over the whole dataset for
    voxel-wise tracer kinetic modeling. Each point represents a voxel within a
    mouse, randomly sampled over \num{50000} voxels. Also for this figure, the fitted
    lines should ideally be overlapping the dashed $y=x$ line}
    \label{fig:scatter_vTCM}
\end{figure}

Downstream implications of AIF estimation accuracy are reflected in kinetic
modeling results shown in Fig~\ref{fig:scatter_vTCM}, where voxel-wise $K_i$
values derived from each model are compared to reference values obtained using
the measured AIF and the Patlak model~\cite{Patlak1983,Patlak1985b}. FC-DLIF
again yields higher agreement with the reference, confirming that improvements
in input function estimation translate to better physiological parameter
estimation.

Results on the unseen tracer data (\fdopa{} and \psma{}) showed decreased performance
for both models, as expected due to tracer-specific uptake characteristics. Additional
scatterplots and analysis are provided in the Appendix~\ref{sec:appendix-additional}
(Figs.~\ref{fig:scatter_AIFs2} and~\ref{fig:scatter_vTCM2}).

\subsection{Qualitative results}

This section evaluates the predictions of the proposed FC-DLIF model. This is
followed by an experiment that shift the PET images in time, to delay tracer
injection start. Then the PET images are truncated, which simulates
simultaneous tracer injection and scan start.

\begin{figure}[t]
    \centering
    \begin{subfigure}[t]{0.32\columnwidth}
        \includegraphics[width=\linewidth,keepaspectratio]{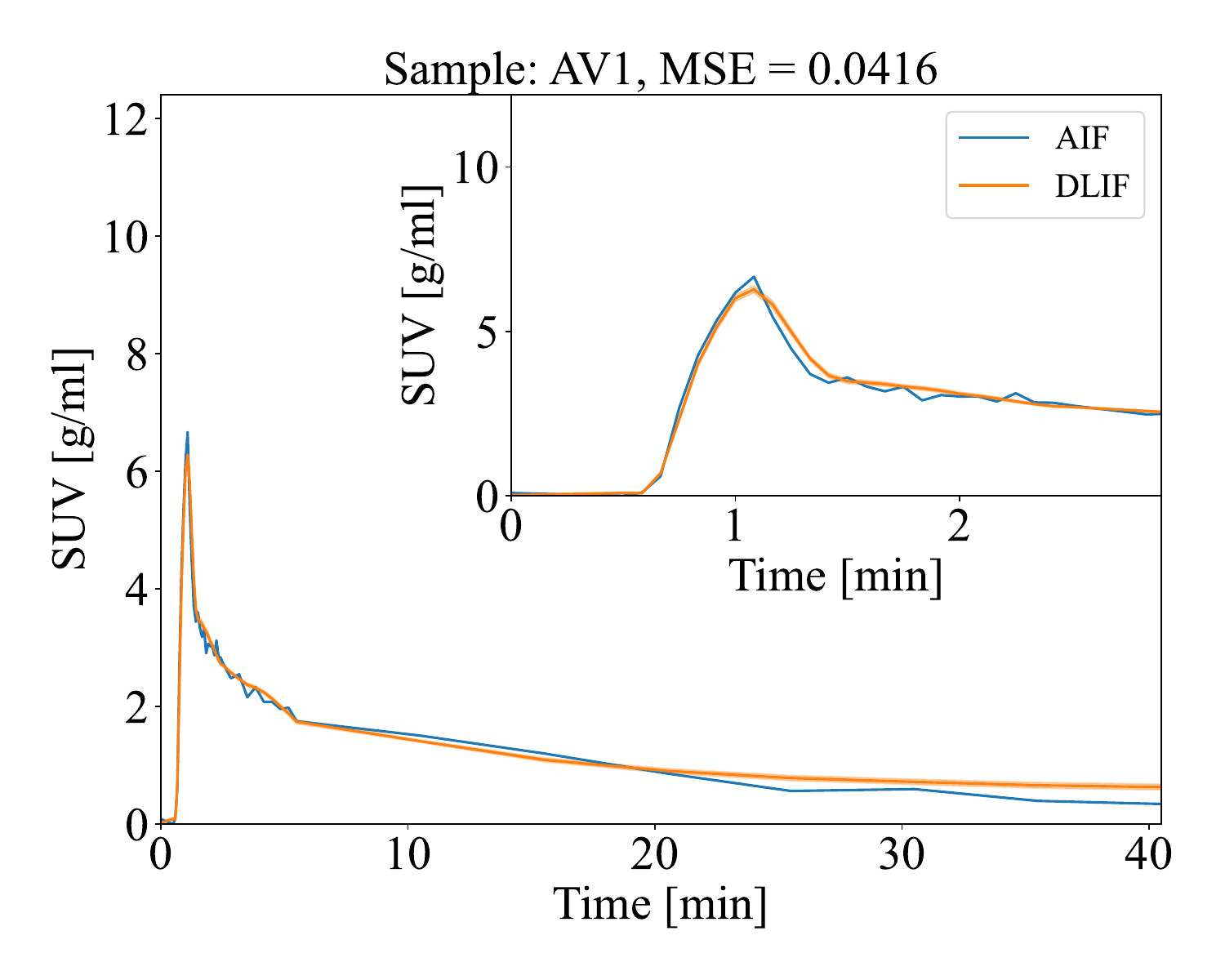}
        \caption{Best sample}
        \label{fig:best_sample}
    \end{subfigure}
    \begin{subfigure}[t]{0.32\columnwidth}
        \includegraphics[width=\linewidth,keepaspectratio]{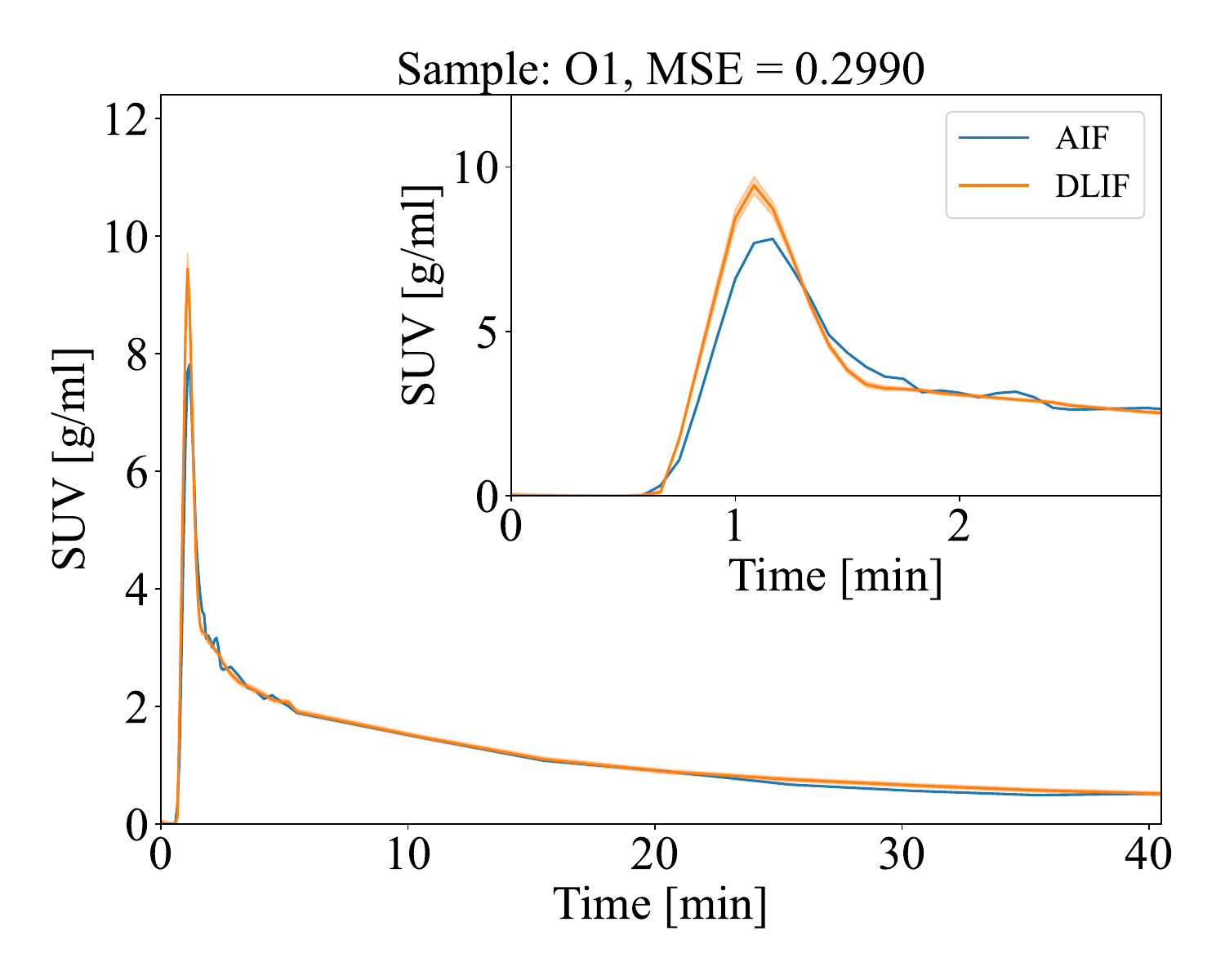}
        \caption{Median sample}
        \label{fig:median_sample}
    \end{subfigure}
    \begin{subfigure}[t]{0.32\columnwidth}
        \includegraphics[width=\linewidth,keepaspectratio]{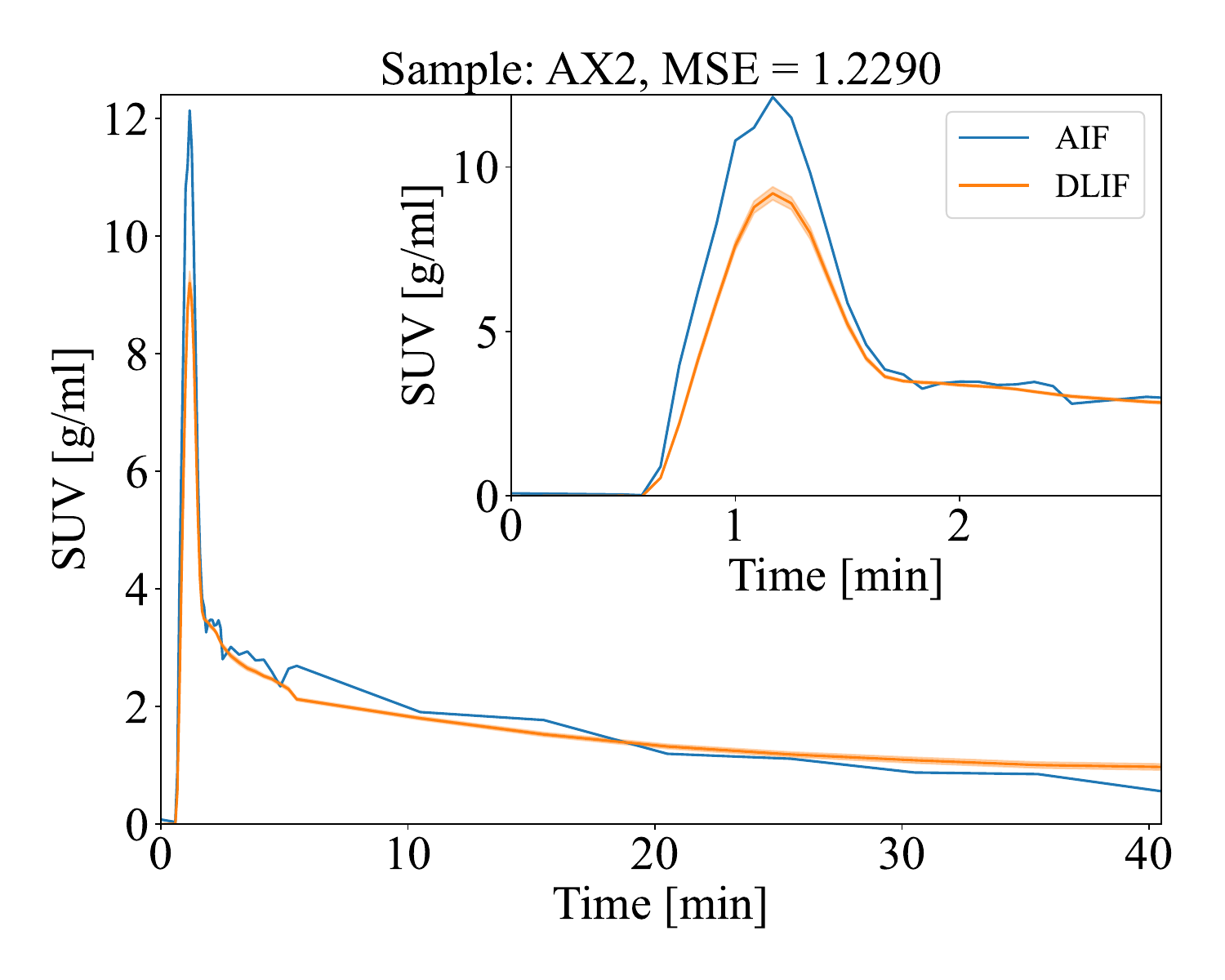}
        \caption{Worst sample}
        \label{fig:worst_sample}
    \end{subfigure}

    \caption[Example input function predictions.]%
    {Examples of input function predicted with FC-DLIF and compared
    against the ground truth AIF. The insets zoom on the first \num{3} minutes
    of the curves, to emphasize the tracer uptake peak. (a) Best sample; (b)
    median sample; (c) worst sample.}
    \label{fig:DLIF_examples}
\end{figure}

Fig.~\ref{fig:DLIF_examples} shows three examples of input functions predicted
by the proposed FC-DLIF model. The mean curve and standard deviation of the
model's predictions are included for the best (Fig.~\ref{fig:best_sample}),
median (Fig.~\ref{fig:median_sample}), and worst (Fig.~\ref{fig:worst_sample})
sample according to the MSE. 

\begin{figure}[t]
    \centering
    \begin{subfigure}[t]{0.48\columnwidth}
        \includegraphics[width=\linewidth,keepaspectratio]{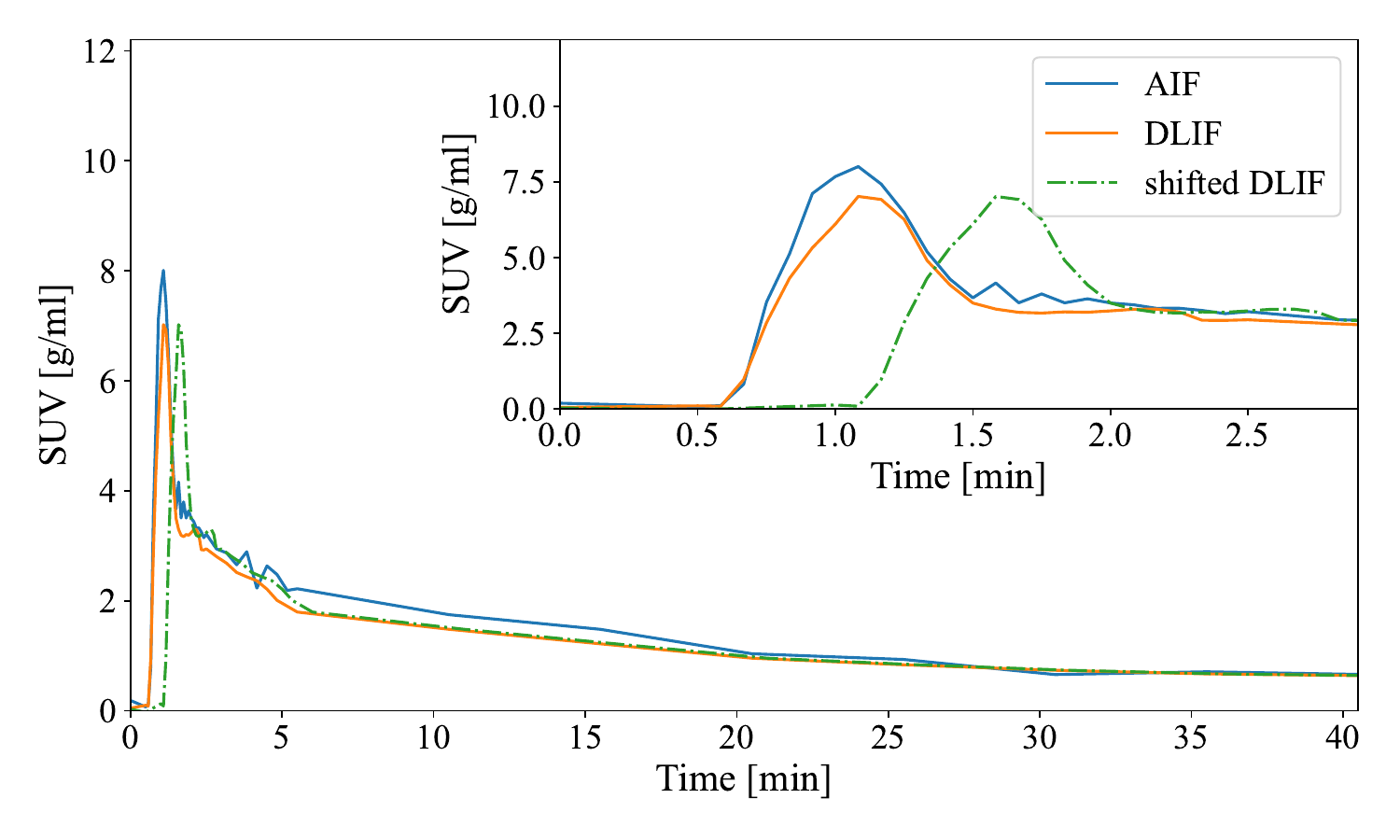}
        \caption{Time-shifted sample}
        \label{fig:shifted}
    \end{subfigure}
    \begin{subfigure}[t]{0.48\columnwidth}
        \includegraphics[width=\linewidth,keepaspectratio]{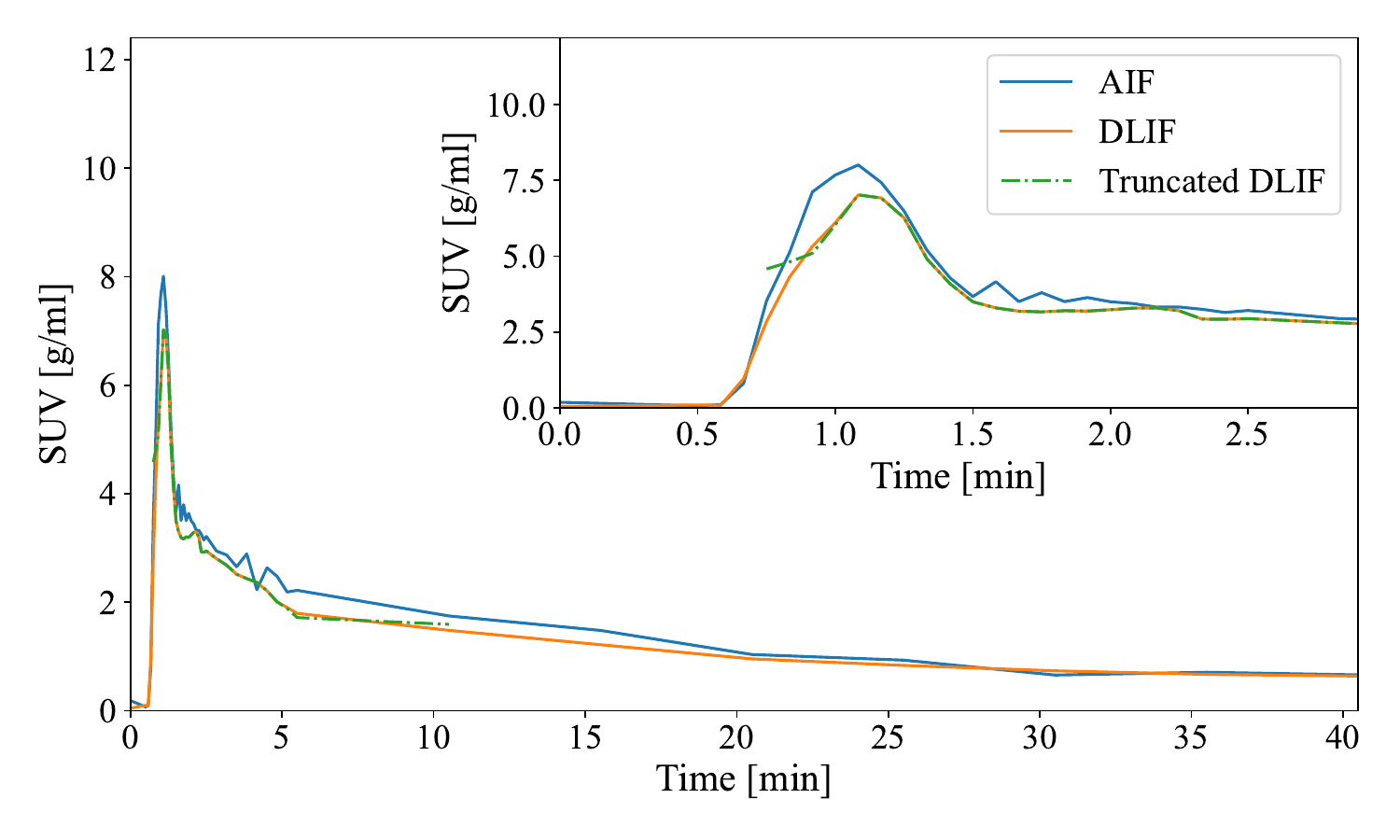}
        \caption{Truncated sample}
        \label{fig:truncated}
    \end{subfigure}

    \caption[Special cases of input functions predicted with FC-DLIF]%
    {Special cases of input function predicted with FC-DLIF. The insets zoom on the
    first \num{3} minutes of the curves: (a) Shifted sample. An additional empty
    \num{30} second frame was added at the beginning of the input time series; (b)
    Truncated sample. The first \num{40} seconds and the last \num{30} minutes of the
    input time series were removed, corresponding to the first \num{4} and last \num{6}
    time steps.}
    \label{fig:versatile}
\end{figure}

Fig.~\ref{fig:versatile} displays the versatility of the proposed model owing to
its fully convolutional design. In particular, the example on the left is the
output of the model when the input data is time-shifted by prepending the first
frame of the dynamic PET image to itself. The example on the right is the result
obtained by truncating the input both at the beginning and at the end of the
time series, removing several time steps.

\begin{figure}[t]
    \centering
    \includegraphics[width=\linewidth]{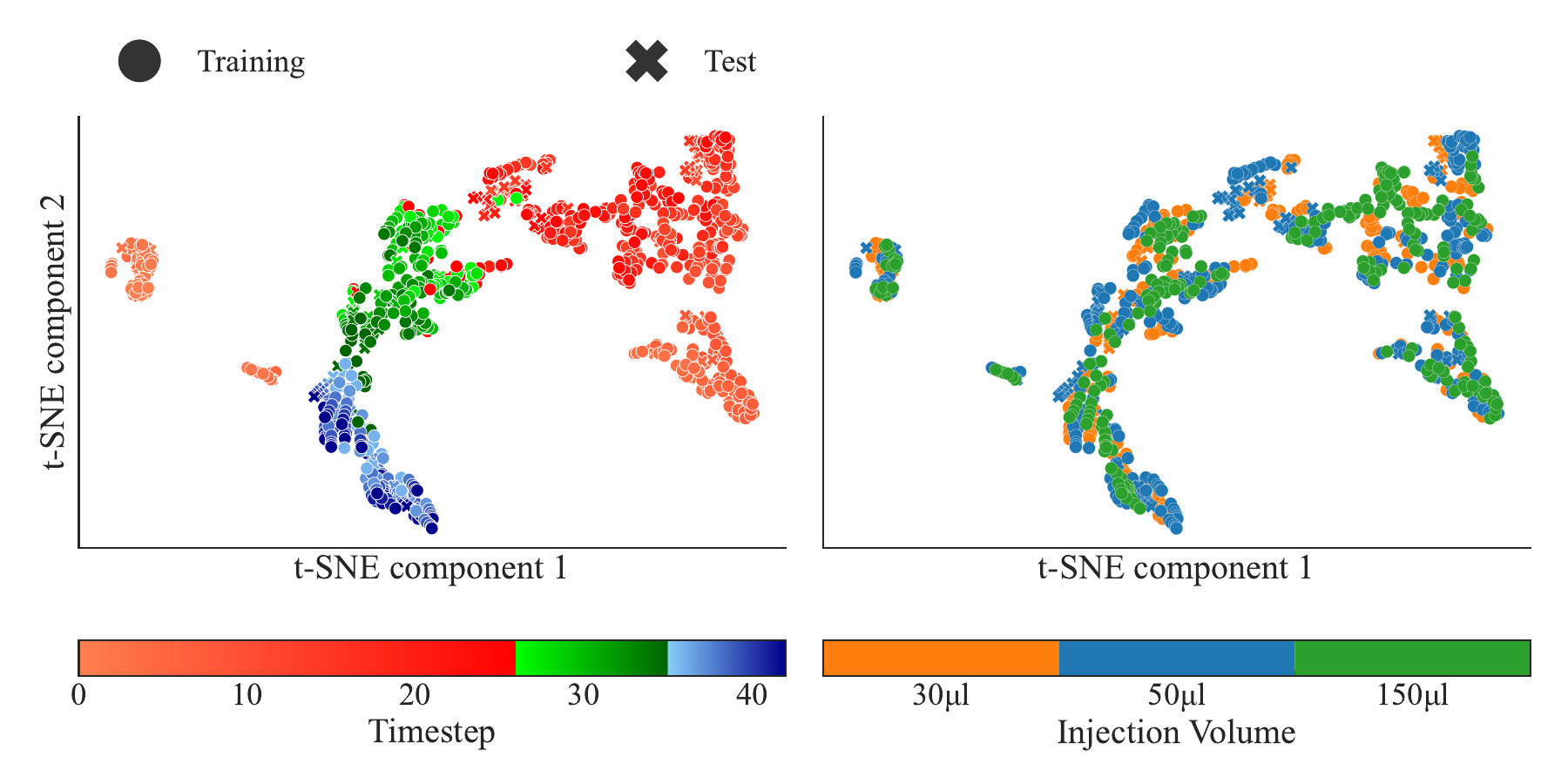}

    \caption[t-SNE visualization of latent features for different injection volumes.]%
    {t-SNE visualization of the SFE-extracted vectors for different
    injection volumes. On the left-hand side, the colormap distinguishes peak
    (red), mid (green), and tail (blue) time steps. To the right, the colormap
    indicates the various mice groups.}
    \label{fig:t-sne-volume}
\end{figure}

\begin{figure}[t]
    \centering
    \includegraphics[width=\linewidth]{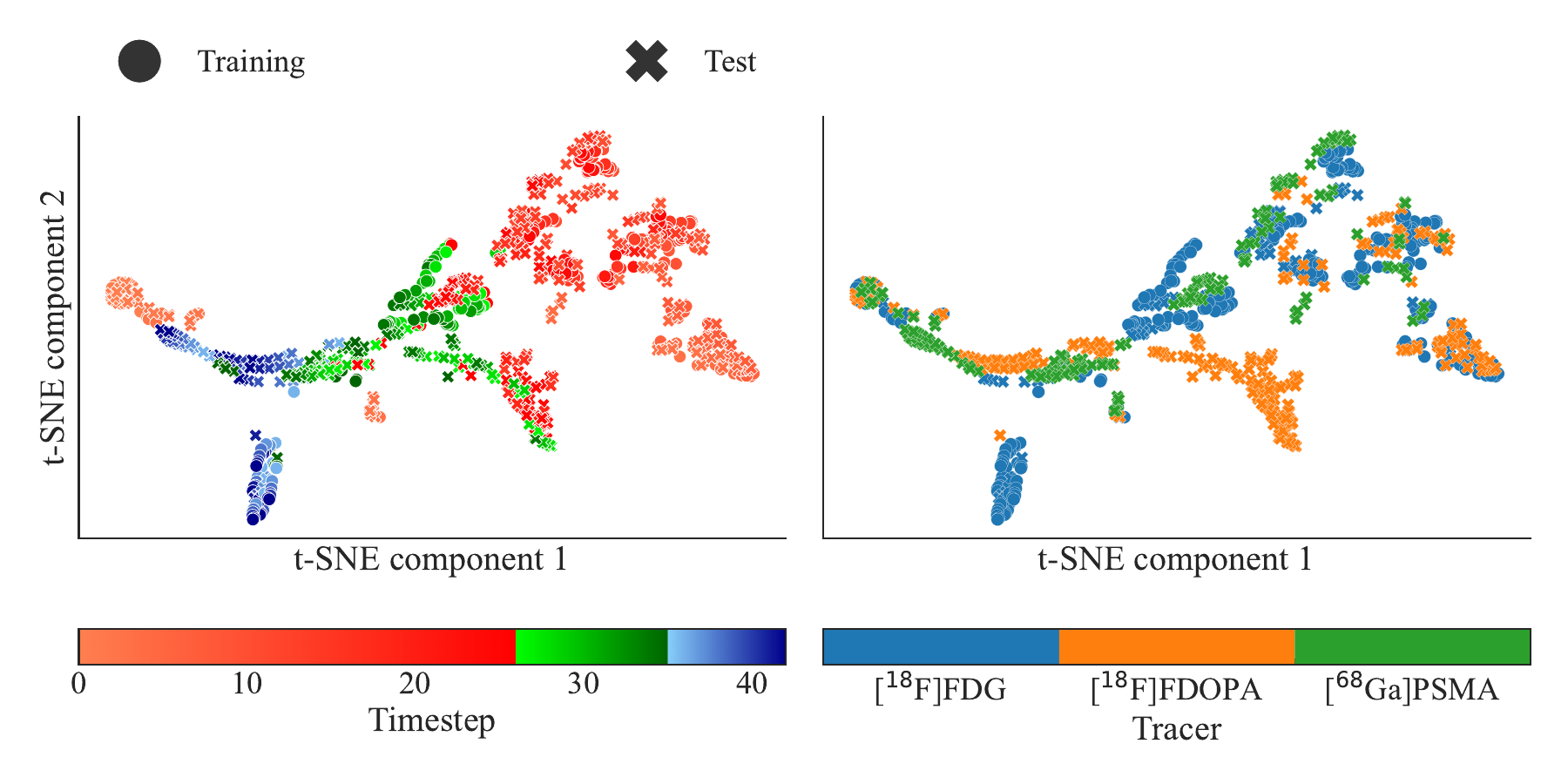}

    \caption[t-SNE visualization of latent features for different injected tracers.]%
    {t-SNE visualization of the SFE-extracted vectors for different
    injected tracers. To the left, the colormap distinguishes peak (red), mid
    (green), and tail (blue) time steps. To the right, the colormap indicates
    the various mice groups.}
    \label{fig:t-sne-tracers}
\end{figure}

Finally, t-SNE is used to visualize the similarities between the
high-dimensional vectors extracted by the SFE for each time step of each mouse,
on both the training and test datasets. Fig.~\ref{fig:t-sne-volume} and
\ref{fig:t-sne-tracers} illustrate the results of the t-SNE dimensionality
reduction for the groups of injected volume and injected tracers respectively.
To the left, the colormap indicates the peak, mid, and tail time steps in shades
of red, green, and blue respectively. On the right hand side, the colormap
indicate certain attributes---the amount of \fdg{} injected in
Fig.~\ref{fig:t-sne-volume}, and what tracer is used in
Fig.~\ref{fig:t-sne-tracers}.

Fig.~\ref{fig:t-sne-volume} shows a clear temporal progression in feature
space. Early frames with low uptake cluster tightly on the left, followed by a
gradual arc through peak and post-distribution frames. The pattern reflects the
biological progression of tracer dynamics and suggests that the SFE captures
consistent spatial characteristics across time.

Likewise, Fig.~\ref{fig:t-sne-tracers} shows the t-SNE projection, from
different tracer groups, that were kept aside during training, comprised of
\fdg{}, \psma{}, and \fdopa{} injected mice. While the early and peak phases of
all tracers appear to follow a similar initial trajectory, clear deviations form
in the later frames. In particular, the steady-state representations of the
\psma{} and \fdopa{} groups form separate clusters, distinct from the \fdg{}
tail-phase points. This indicates that the model's spatial encoder captures
consistent tracer-specific signatures not seen during training. Additionally,
the \fdopa{} group also diverges around the late peak phase, forming a distinct
cluster that may explain the model’s reduced accuracy for this group. These
observations align with the systematic under- and overestimation patterns seen
in the early and late phases of predicted AIFs for these out-of-distribution
tracers.

\section{Discussion}

This study introduces FC-DLIF, a robust and flexible deep learning model for the
non-invasive estimation of AIF directly from dynamic PET imaging data. Compared
to existing methods, FC-DLIF demonstrates superior accuracy, reduced bias, and
improved robustness to noise and input variability, while maintaining
flexibility with respect to temporal shifts and varying scan lengths. The fully
convolutional design enables the model to generalize across different imaging
protocols without the need for rigid preprocessing or fixed input dimensions.

\subsection{Quantitative analysis}

\paragraph{Fully convolutional design leads to consistent improvements}

Across all quantitative evaluations, FC-DLIF demonstrated superior accuracy
compared to the baseline model. While both methods use 3D convolutional
backbones to extract spatial features, the fully convolutional design of
FC-DLIF, that comprises separate spatial and temporal feature extractors,
appears to contribute to more accurate AIF predictions. In particular, each time
frame is processed independently through the same spatial filters in the SFE,
potentially reducing time-point-specific biases during feature extraction. The
subsequent TFE, built with 1D convolutions (Fig.~\ref{fig:model}), is then able
to learn temporal dependencies over a time-series of uniformly extracted spatial
representations.

Although the precise source of performance gain is difficult to isolate, the two-step
network inherently enforces a type of implicit regularization. This separation of
concerns allows each subnetwork to specialize; spatial encoding in the SFE; and temporal
modeling in the TFE, potentially introducing a form of structural regularization. This
modular architecture may help the network learn more generalized temporal patterns in
tracer dynamics, leading to improved alignment with the true AIF. The benefits are
clearly reflected in the lower MSE distribution (Fig.~\ref{fig:mse-boxplot}), being less
affected by a noisy input signal (Fig.~\ref{fig:noise}), and a tighter correspondence
between predicted and true AIF values (Fig.~\ref{fig:scatter_AIFs}).

\paragraph{Improved downstream kinetic modeling}

The improvements in AIF prediction also translate into better physiological parameter
estimation, as evidenced by the voxel-wise $K_i$ comparisons
(Fig.~\ref{fig:scatter_vTCM}), and region-wise comparisons (Fig.~\ref{fig:Ki myocardium}
and Fig.~\ref{fig:Ki brain}). Accurate AIF estimation is critical for computing reliable
kinetic parameters using models such as Patlak~\cite{Patlak1983,Patlak1985b} and
compartment models~\cite{sokoloff1977kineticmodeling}, and FC-DLIF's enhanced
performance further supports its potential to replace invasive blood sampling without
compromising modeling accuracy.

\paragraph{Generalization challenges with unseen tracers}

When applied to a dataset using tracers not seen during training (\fdopa{} and \psma{}
tracers), both models exhibited reduced performance. This result is consistent with the
expectation that different tracers produce distinct uptake patterns and temporal
kinetics, making generalization difficult without exposure during training. Future work
may explore strategies for improving generalization across tracers, such as domain
adaptation~\cite{transferlearning2021zhuang, domainadaptation2021Farahani}, tracer-aware
conditioning, or semi-supervised strategies~\cite{chapelle2009SSL}.

\subsection{Qualitative analysis}
\label{sec:qualitative analysis}

\paragraph{Understanding prediction characteristics}

To further gain insight into the model's behavior, the AIF predictions from FC-DLIF were
examined over a range of samples (Fig.~\ref{fig:DLIF_examples}). Even in the
worst-performing example, the model captures the general shape of the AIF, particularly
during the post-distribution phase. This aligns with the scatter-based results from
Fig.~\ref{fig:scatter_AIFs}, where the largest errors are observed around the peak
uptake frames, while the steady-state phase shows more consistent alignment with ground
truth. This discrepancy likely stems from the temporal characteristics of the input data
itself: the uptake phase is captured in only a few frames, and exhibits high inter-frame
variability, making it difficult to learn reliably. In contrast, the tail of the curve,
corresponding to tracer equilibrium, spans more frames and features more consistent
signal characteristics, yielding more stable predictions (as seen in Fig.~\ref{fig:curve
phases}).

Importantly, for downstream kinetic modeling tasks such as Patlak
analysis~\cite{Patlak1983,Patlak1985b}, the steady-state segment of the AIF is the only
part used to derive macro-parameters like $K_i$. This suggests that even if the peak
predictions are imperfect, FC-DLIF's accurate modeling of the linear phase still enables
reliable kinetic quantification, when using the Patlak model.

\paragraph{Robustness to shifts and truncation}

One of the practical advantages of FC-DLIF is its architectural flexibility, which is
demonstrated by the model's robustness to time shifts and partial input sequences, a
trait which has not been seen in other similar works. When the input time series was
artificially shifted or truncated, the model continued to produce consistent AIF
predictions (Fig.~\ref{fig:versatile}). This behavior stems directly from the model’s
fully convolutional structure, which applies the same spatial and temporal filters
regardless of absolute time index or input length, emphasizing its practicality in
real-world settings, where scan durations and start times may vary between protocols or
subjects.

\paragraph{What the model learns from spatial features}

To better understand the internal representations learned by FC-DLIF, the spatial
features extracted by the SFE were visualized using t-SNE~\cite{van2008visualizing}.
Across all mice and time points, the t-SNE component projection reveals a coherent
temporal progression: from early frames with low uptake, through peak uptake, to the
post-distribution tail. This suggests that the SFE successfully encodes temporally
meaningful spatial information, even though it processes each frame independently.
Notably, these feature clusters were consistent across both training and test sets, with
no clear separation, indicating good generalization and successfully not overfitting.
Furthermore, the internal feature space showed no meaningful correlation with injected
volume, suggesting that the SFE learned to abstract away from trivial scaling
differences in tracer volume, which is an encouraging sign of biological relevance and
robustness (Fig.~\ref{fig:t-sne-volume}).

\paragraph{Out-of-distribution tracers reveal biological misalignment}

Despite showing reasonable results and potential for improvements, t-SNE projections of
mice injected with unseen tracers (\psma{} and \fdopa{}) revealed key deviations. While
their early and peak-phase representations partially aligned with the \fdg{} training
set, their tail-phase features diverged and formed distinct clusters. This reflects
fundamental differences in tracer kinetics and binding
properties~\cite{fdg-fdopa-psma2021duclos}. Correspondingly, FC-DLIF struggled to
predict accurate AIFs for these tracers, typically underestimating peak intensity and
overestimating post-distribution activity. This mismatch to some extent mirrors the
temporal drift observed in the feature space, and suggests that the model’s learned
temporal filters in the TFE are not readily transferable across tracer types without
retraining.

\paragraph{Spatial dimensions limitations}

While the proposed method accepts input data of varying temporal dimensions, a
pretrained model constrains future studies to preprocess their spatial dimensions to
match that of the data it was trained on. Since transformations and interpolations over
spatial dimensions are common and easier to visualize than temporal ones, this is not
seen as a large hurdle. 

\paragraph{Toward generalization through transfer learning}

Despite these limitations, the structure of the learned feature space provides an
opportunity for targeted adaptation. Since the SFE appears to produce meaningful spatial
embeddings even for unseen tracers, it may be possible to freeze this component and
retrain only the temporal extractor (TFE) on a small amount of new tracer data, to map
the distinct clusters back onto the common feature manifold. This strategy aligns with
principles from transfer learning and domain
adaptation~\cite{transferlearning2021zhuang,domainadaptation2021Farahani}, and could
substantially reduce the data requirements for generalizing FC-DLIF to new tracers.
Given that most variation across tracers lies in their temporal behavior, this modular
approach, reusing spatial filters while tailoring the temporal decoder, offers a
promising path forward.

\paragraph{Compact design for efficient deployment}

A final yet key advantage of FC-DLIF is its compact size. The model end-to-end processes
full 4D PET volumes yet comprises only \num{90 124} trainable parameters (352 KB). This
efficiency arises from its internal structure, where each 3D time frame is processed
independently using shared spatial filters in the SFE, followed by a lightweight 1D
convolutional TFE consisting of 4 layers. For comparison, the input images are
approximately 71 MB (64-bit floating point) each, making the model over 200 times
smaller than a single input scan. Despite its modest footprint, which is comparable to
the \num{92 104} parameter (360 KB) baseline model, FC-DLIF achieves superior predictive
performance, making it an attractive candidate for real-time or embedded applications.

\section{Conclusion}

This study introduces a non-invasive, deep learning-based input function predictor, that
provides a flexible method that is invariant to time-shifting and truncation of the
sequence length. Altogether, FC-DLIF represents a promising tool for preclinical PET
studies, removing the need for invasive blood sampling and enabling more efficient and
scalable experimental designs.

\paragraph{Acknowledgments}

This work is funded through \emph{Visual Intelligence}, a Center for Research-based
Innovation funded by the Research Council of Norway, grant no. 309439 and 303514. It was
further supported by Tromsø Research Foundation (19\_PET-NUKL) through the 180$^\circ$N
Norwegian Nuclear Medicine Consortium and UiT Innovation funds (UiT Talent).

\bibliography{bibliography}

\appendix

\section{Data acquisition and processing}
\label{sec:appendix-data}

\subsection{Animal experiments}

The imaging data of mice used to train and evaluate the FC-DLIF model were acquired at
UiT The Arctic University of Norway (UiT). All animal experiments were approved by the
Norwegian Food Safety Authority; FOTS id 29689. In total, \num{80} healthy female mice
from three different strains were included in this study: BALB/cJRj (N = \num{55}),
C57BL/6JRj (N = \num{8}) and Balb/cAnNCrl (N = \num{7}), where BALB/cJRj made up the
main group of \num{70} samples. The remaining \num{10} were set aside for evaluating the
effects of the tracers \fdopa{} (N = \num{6}) and \psma{} (N = \num{4}). The mice were
\numrange{7}{8} weeks of age upon arrival to the UiT animal facility, and fed ad libitum
a standard rodent diet.

\subsection{Arterial and venous cannulation during image acquisition}

At the time of PET/CT imaging, the mouse age was \SI{13.0 +- 0.9}{\weeks}, with
a corresponding weight of \SI{22.5 +- 0.5}{\gram}. The study used three different
radiotracers: \fdg{}, \fdopa{}, and \psma{}. The mice receiving \fdg{}
(\num{70} animals) were fasted for \SI{3.4 +- 0.2}{\hours} prior to injection,
whereas the remaining animals---given \fdopa{} and \psma{}---were not fasted.
The mice were anesthetized prior to the scan, weighed, and placed on a heated plate
(\SI{38}{\degreeCelsius}) while receiving oxygen through a mask. A venous catheter was
inserted into the tail vein of each mouse for radiotracer injection. The blood glucose
during venous cannulation was measured to \SI{6.4 +- 0.3}{\milli\mol\per\liter}. An
incision in the neck enabled surgical cannulation of the carotid artery, allowing blood
to be routed through a radiation detector at a withdrawal rate of
\SI{105.6 +- 2.8}{\micro\liter\per\minute}. This setup facilitated concurrent
measurements of whole blood activity during the PET scan with a temporal resolution of
\SI{1}{\second}. To enable continuous arterial line measurements, an arterial--venous
shunt was established, forming a closed loop between arterial sampling and venous
reinjection. Blood flowed sequentially from the arterial sampling line, radiation
detector, a peristaltic pump, and a Y-connector---enabling intravenous radiotracer
injection---before being reinfused via the venous injector, thereby preventing excessive
blood loss.

The PET/CT imaging was performed using a \qty{45.5}{\minute} listmode scan on a
Triumph$^{\mathrm{TM}}$ LabPET-8$^{\mathrm{TM}}$ small animal PET/CT scanner (TriFoil
Imaging Inc., Chatsworth, CA, USA) while a sensor monitored the respiration rate. The
mice were injected with \qty{16.2 +- 0.7}{\mega\becquerel} using an automated injection
pump, started \SI{30}{\second} after scanning was initiated. CT imaging was performed
after PET scanning to correct for attenuation and scatter. While still under deep
anesthesia after scanning, the mice were euthanized using cervical dislocation. Scanner
sensitivity was monitored through daily phantom calibrations.

\subsection{Calibration and AIF processing}

To allow for delay correction of the manual blood sample measurements in the
arteriovenous shunt, the time delay between the radiation detector and the Y-connector
was measured during the first pass of arterial blood to \SI{25.1 +- 0.7}{\second}. The
continuous line radiation detector was calibrated with three manual blood samples taken
in a late stage or post-scan by measuring blood dripped from the arterial catheter over
\SI{30}{\second}. This also enabled measurement of the actual arterial withdrawal rate
during each mouse scan.

A calibration factor for the continuous arterial line measurements was derived as the
ratio of the average signal from the continuous line measurements and the manual blood
samples collected during the same \SI{30}{\second} blood sampling at each time point,
corrected for delay. Calibration factors outside three scaled median absolute deviations
from the median factors were considered outliers and discarded~\cite{Hubert2008}. An
overall calibration factor was determined as the average factor from the included blood
sample factors. The arterial input function (AIF) for each mouse was obtained by scaling
the continuous line measurement data by this average calibration factor.

\subsection{Image reconstruction and processing}

The PET images were reconstructed into \num{42} time frames
(\qtylist[parse-numbers=false]{\numproduct{1x30};\numproduct{24x5};\numproduct{9x20};\numproduct{8x300}}{\second})
using a three-dimensional maximum-likelihood estimator algorithm with \num{50} iterations.
Corrections for detector efficiency, radioactive decay, random coincidences, dead time,
attenuation, and scatter were applied. Each time frame had an image matrix size of
\numproduct{128x92x92} voxels. The voxels were converted from units of counts per second
into units of \unit{\mega\becquerel\per\milli\liter} using the average counts inside a
\SI{14}{\milli\liter} homogeneous image region of a daily phantom scan. Subsequently, the
voxels were normalized into standardized uptake value (SUV) [\unit{\gram\per\milli\liter}]
\cite{Keyes1995a}.

\begin{figure}[t]
    \centering
    \includegraphics[width=\linewidth]{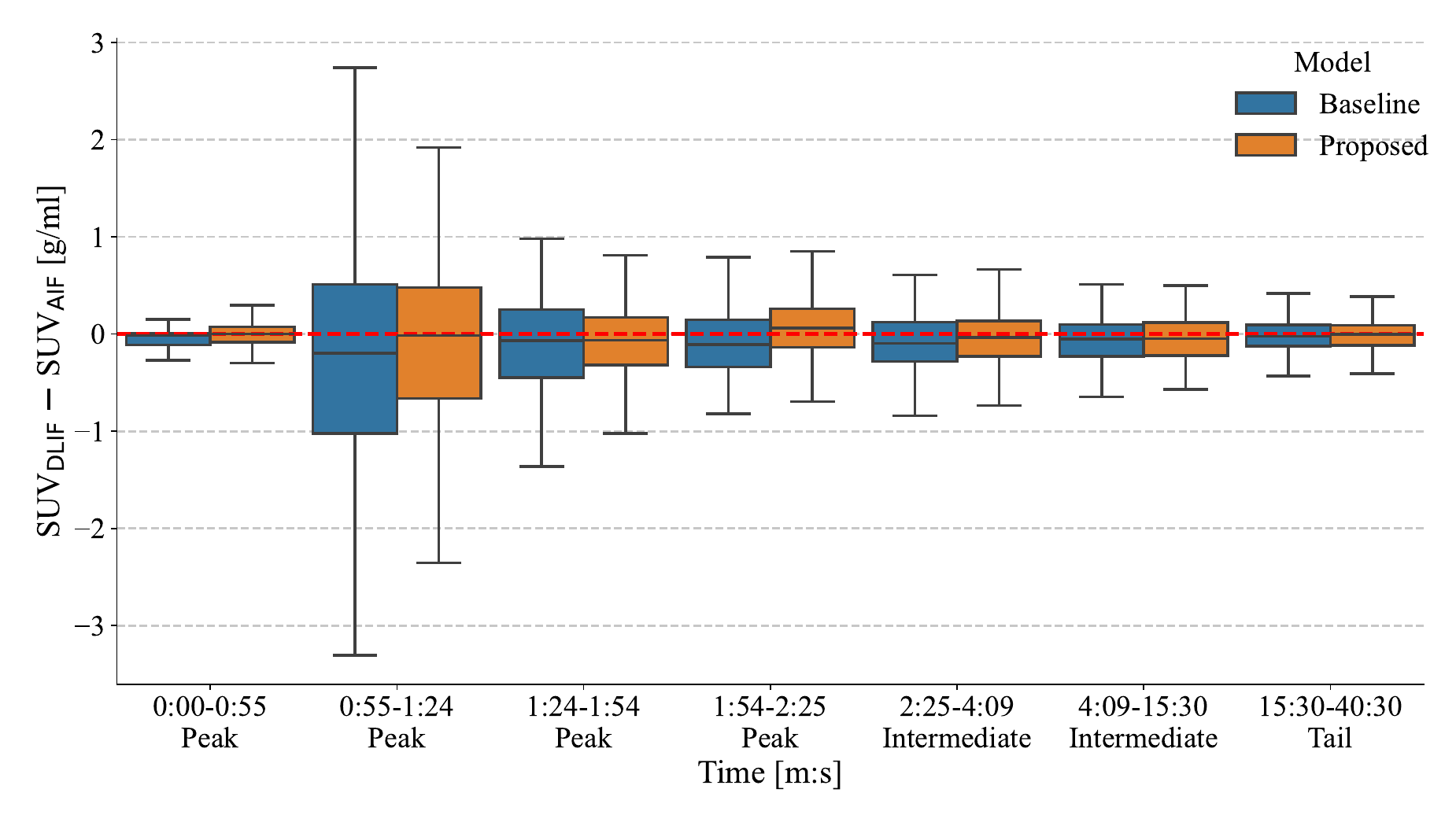}

    \caption[Distribution of prediction errors in different phases of input function.]%
    {Prediction error distribution during different phases of the imaging for
    the baseline DLIF~\cite{kuttner2024deep} and the proposed FC-DLIF.}
    \label{fig:curve phases}
\end{figure}

\section{Additional results}
\label{sec:appendix-additional}

\subsection{Detailed results for individual imaging phases}
\label{sec:appendix-curve-segments}

Figure~\ref{fig:curve phases} visualizes the distribution of error between the predicted
input function from the baseline model~\cite{kuttner2024deep} and FC-DLIF compared with
the AIF, as a function of time since the start of tracer injection. A predominantly
negative distribution in the plot would indicate the model underestimating the AIF.

\subsection{Results on unseen tracers dataset}
\label{sec:appendix-external-dataset}

\begin{figure}[t]
    \centering
    \begin{subfigure}[t]{0.49\linewidth}
        \includegraphics[width=\linewidth,keepaspectratio]{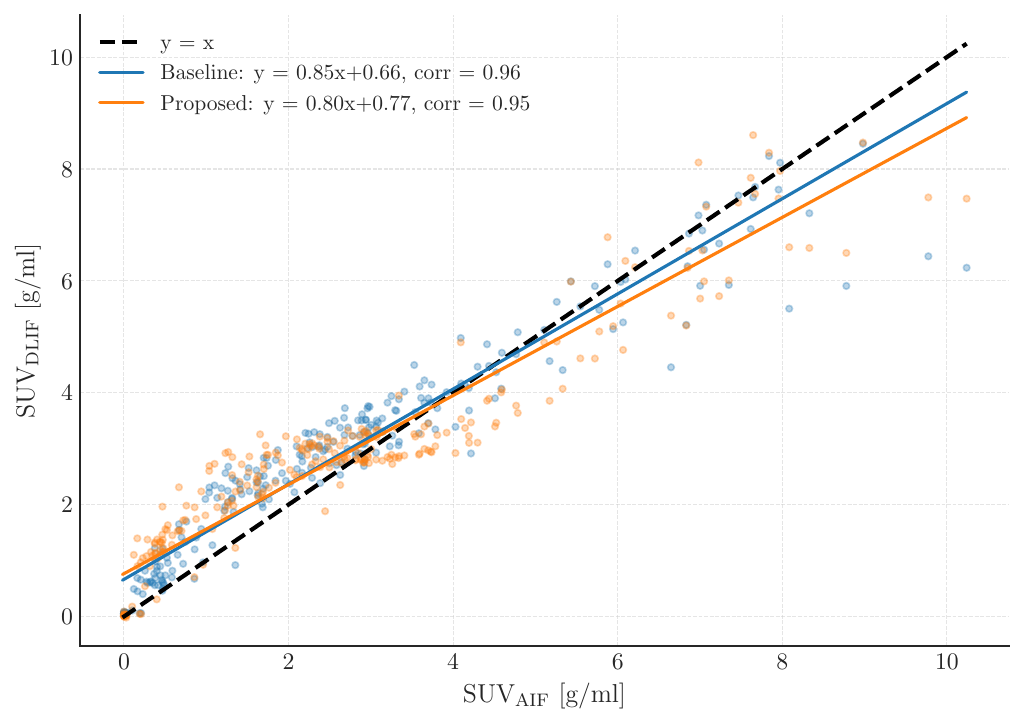}
        \caption{FDOPA}
        \label{fig:scatter_AIFs2 FDOPA}
    \end{subfigure}
    \begin{subfigure}[t]{0.49\linewidth}
        \includegraphics[width=\linewidth,keepaspectratio]{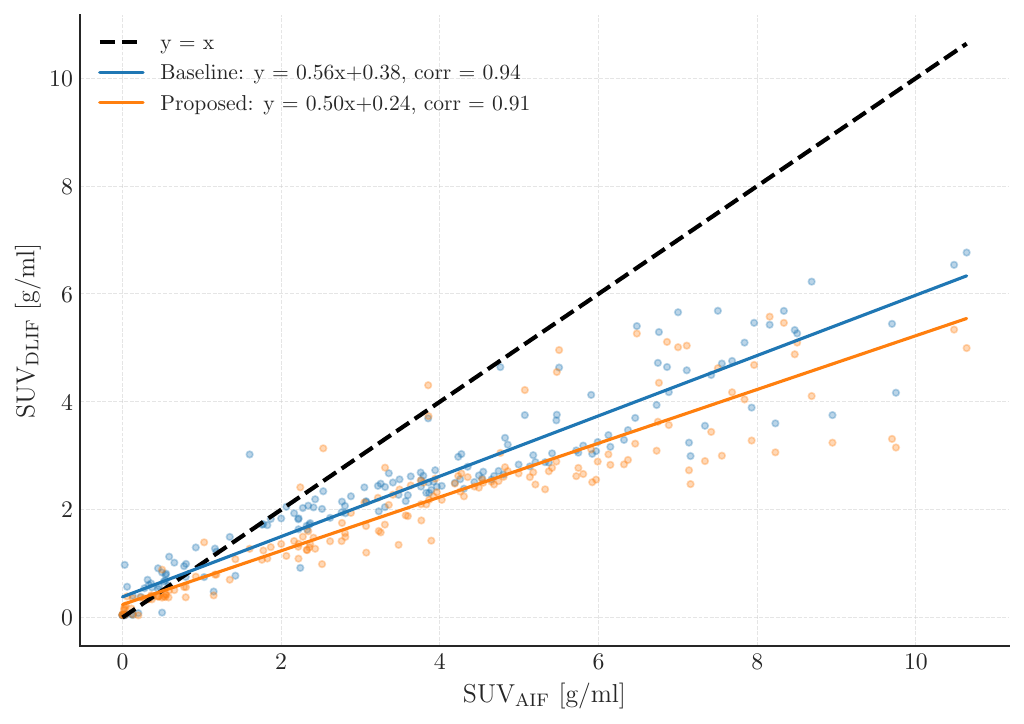}
        \caption{PSMA}
        \label{fig:scatter_AIFs2 PSMA}
    \end{subfigure}

    \caption[Scatterplots of performance on unseen tracers.]%
    {Scatterplots summarizing the results over the unseen tracers dataset for
    the arterial input function estimation. The color represents SUV predictions for each
    model.}
    \label{fig:scatter_AIFs2}
\end{figure}

Figure~\ref{fig:scatter_AIFs2} compares the baseline~\cite{kuttner2024deep} and the
proposed FC-DLIF models' SUV predictions with the measured AIF. The lines are computed
as the orthogonal regression line fitted to the point distributions of each respective
color. A perfect predictor would have SUV points distributed across the black, dashed,
$y=x$ line.

\begin{figure}[t]
    \centering
    \begin{subfigure}[t]{0.49\linewidth}
        \includegraphics[width=\linewidth,keepaspectratio]{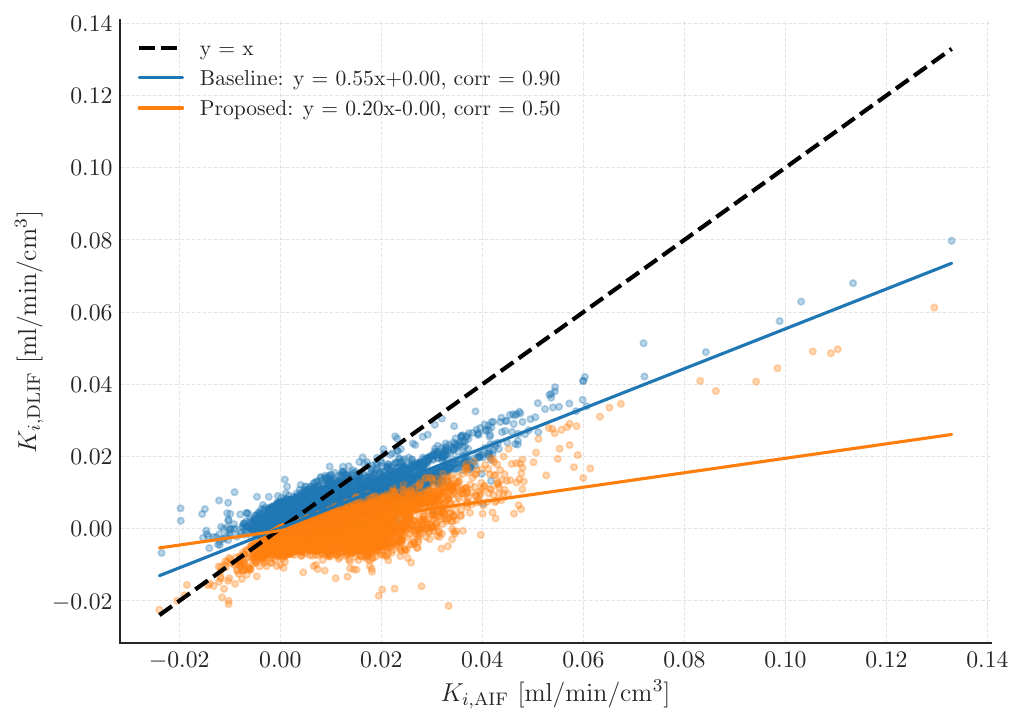}
        \caption{FDOPA}
        \label{fig:scatter_vTCMs2 FDOPA}
    \end{subfigure}
    \begin{subfigure}[t]{0.49\linewidth}
        \includegraphics[width=\linewidth,keepaspectratio]{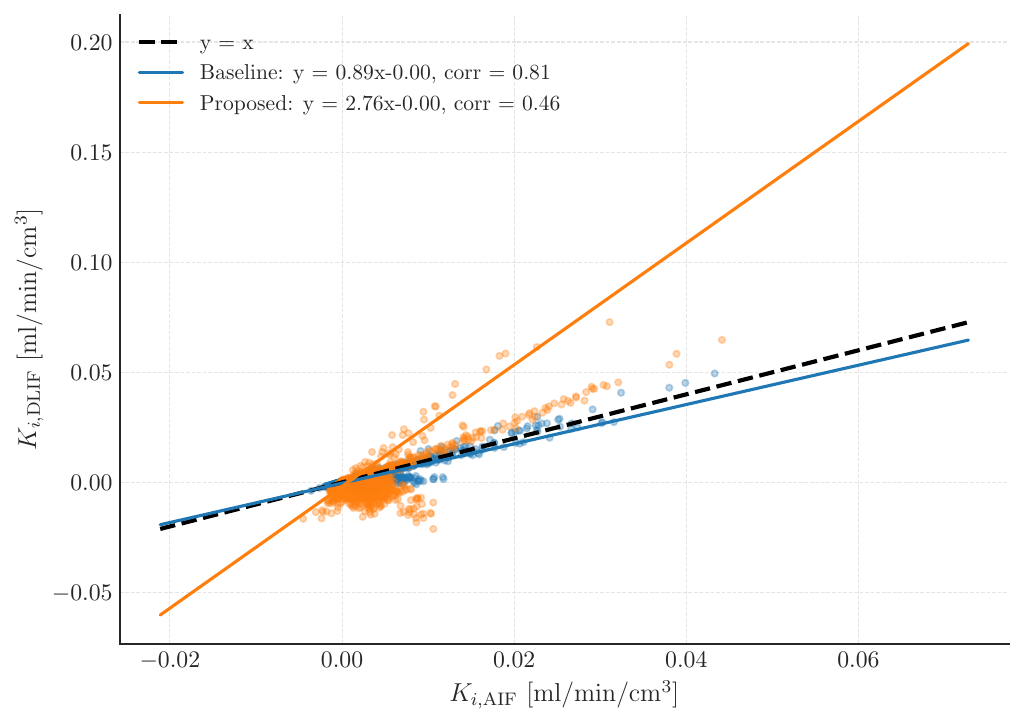}
        \caption{PSMA}
        \label{fig:scatter_vTCMs2 PSMA}
    \end{subfigure}

    \caption[Scatterplots of kinetic parameters on unseen tracers.]%
    {Scatterplots summarizing the results over the unseen tracers dataset for
    voxel-wise tracer kinetic modeling. Colors indicate which model provided each
    coefficient for a random subset of \num{50 000} voxels.}
    \label{fig:scatter_vTCM2}
\end{figure}

Figure~\ref{fig:scatter_vTCM2} uses the predicted input function from either model in a
Patlak graphical analysis~\cite{Patlak1983,Patlak1985b} to derive the kinetic parameters
$K_i$.

\begin{figure}[t]
    \centering
    \begin{subfigure}[t]{0.32\linewidth}
        \includegraphics[width=\linewidth,keepaspectratio]{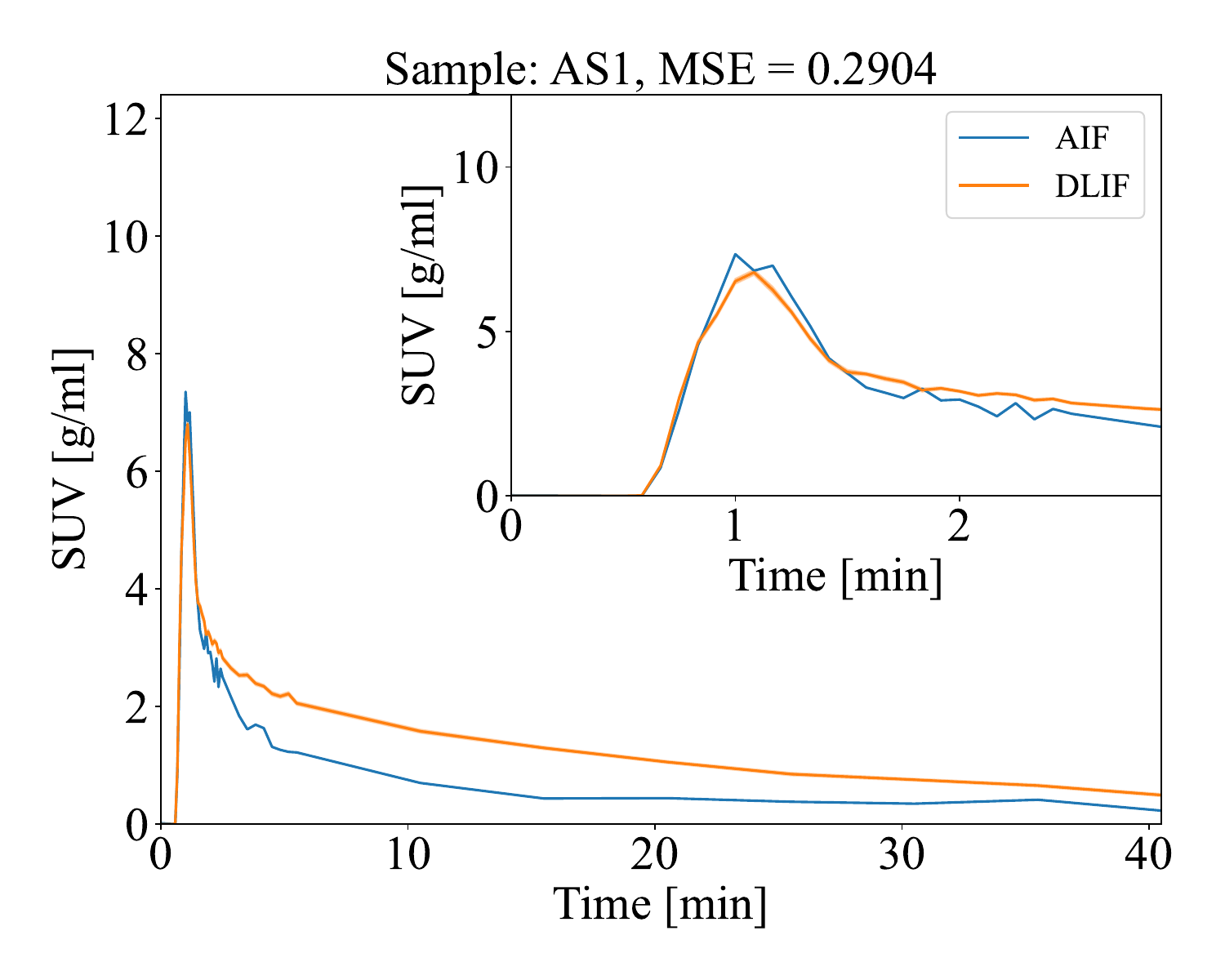}
        \caption{Best sample}
        \label{fig:best_sample2}
    \end{subfigure}
    \begin{subfigure}[t]{0.32\linewidth}
        \includegraphics[width=\linewidth,keepaspectratio]{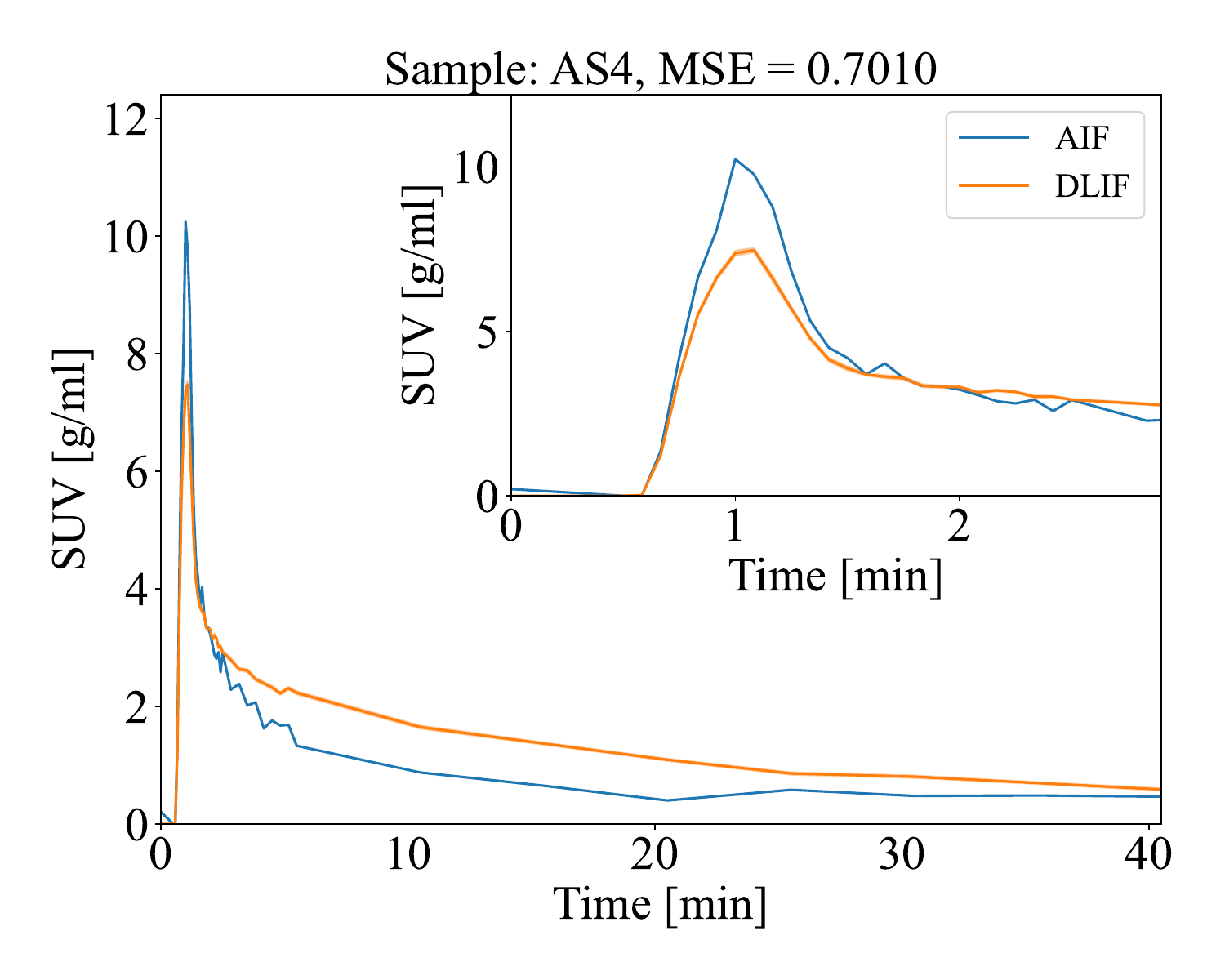}
        \caption{Median sample}
        \label{fig:median_sample2}
    \end{subfigure}
    \begin{subfigure}[t]{0.32\linewidth}
        \includegraphics[width=\linewidth,keepaspectratio]{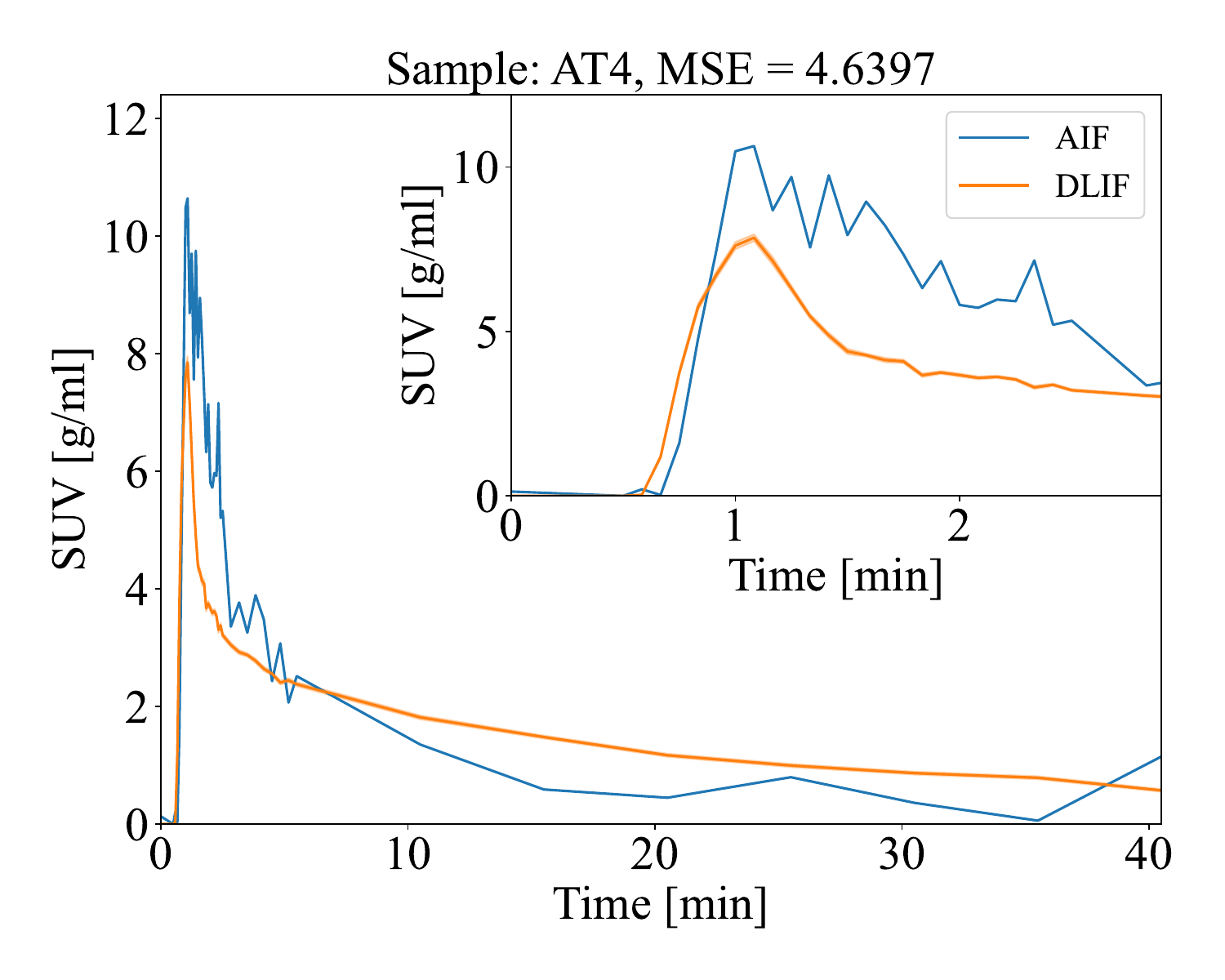}
        \caption{Worst sample}
        \label{fig:worst_sample2}
    \end{subfigure}

    \caption[Examples of input function predictions with FC-DLIF on unseen tracers]%
    {Examples of input function predictions with FC-DLIF compared against the ground
    truth on unseen tracers. Insets zoom on the first \num{3} minutes of the curves,
    when the input function peak occurs. (a) Best sample (\fdopa{}); (b) median sample
    (\fdopa{}); (c) worst sample (\psma{}).}
    \label{fig:DLIF_examples2}
\end{figure}

Figure~\ref{fig:DLIF_examples2} shows examples of input functions predicted by the
proposed FC-DLIF model. The mean curve and the standard deviation over the \num{10} runs
are depicted for the best (left), median (middle), and worst (right) sample according to
the mean squared error (MSE).

\subsection{Detailed kinetic modeling results}
\label{sec:appendix-kinetic}

\begin{figure}[t]
    \centering
    \begin{subfigure}[t]{0.32\linewidth}
        \includegraphics[width=\linewidth,keepaspectratio]{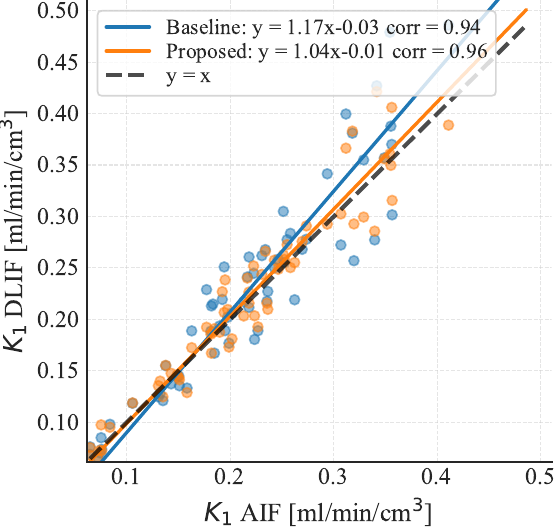}
        \caption{$K_1$}
        \label{fig:K1 myocardium}
    \end{subfigure}
    \begin{subfigure}[t]{0.32\linewidth}
        \includegraphics[width=\linewidth,keepaspectratio]{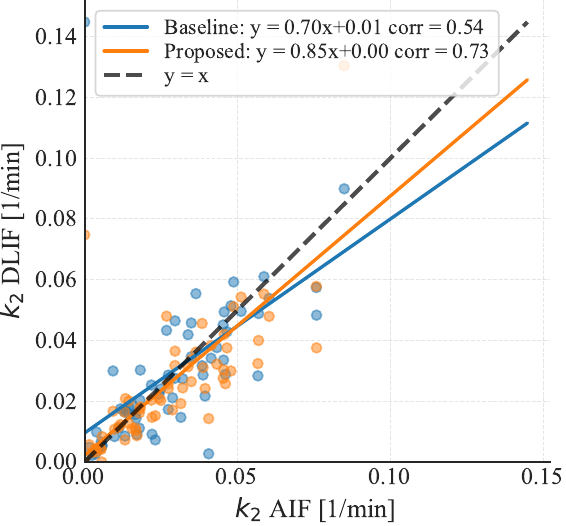}
        \caption{$k_2$}
        \label{fig:k2 myocardium}
    \end{subfigure}
    \begin{subfigure}[t]{0.32\linewidth}
        \includegraphics[width=\linewidth,keepaspectratio]{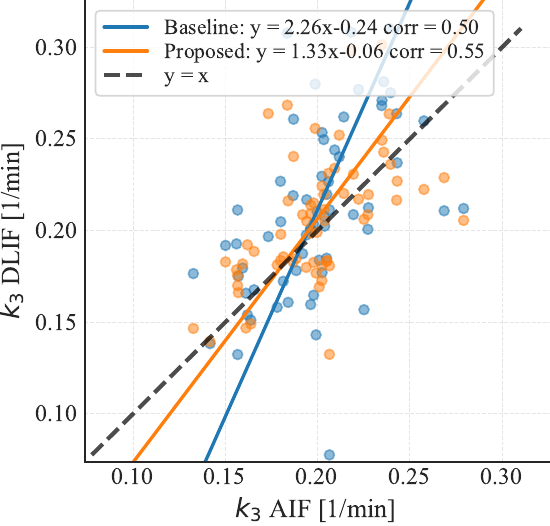}
        \caption{$k_3$}
        \label{fig:k3 myocardium}
    \end{subfigure}
    \begin{subfigure}[t]{0.33\linewidth}
        \includegraphics[width=\linewidth,keepaspectratio]{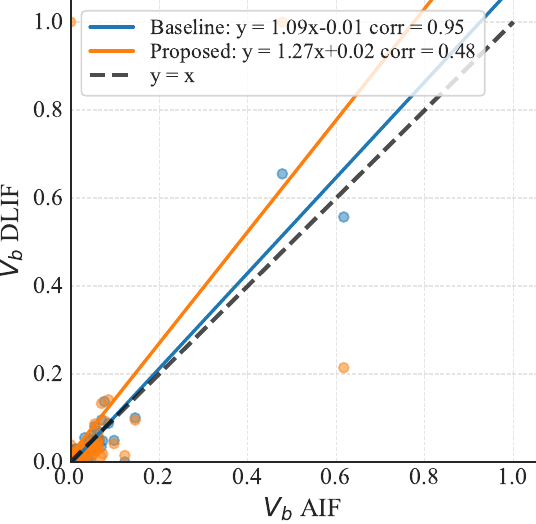}
        \caption{$V_b$}
        \label{fig:Vb myocardium}
    \end{subfigure}
    \begin{subfigure}[t]{0.33\linewidth}
        \includegraphics[width=\linewidth,keepaspectratio]{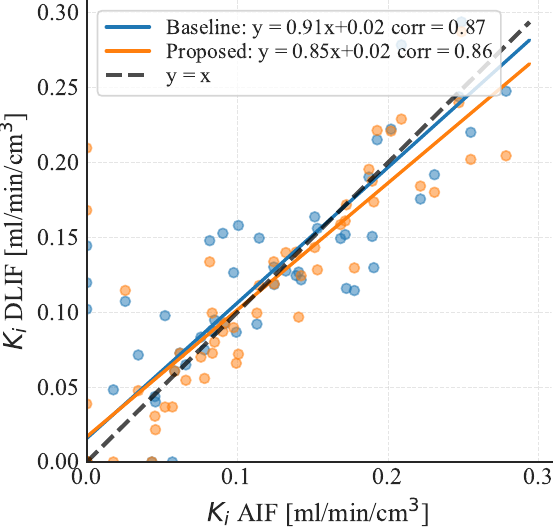}
        \caption{$K_i$}
        \label{fig:Ki myocardium}
    \end{subfigure}

    \caption[Estimated myocardium kinetic parameters using irreversible two-tissue
    compartment model.]%
    {Scatterplots summarizing kinetic parameters estimated using an irreversible
    two-tissue compartment model in the myocardium region. Points represent each region
    in each mouse sample, colored by the model producing the coefficient. From top left
    to bottom right: $K_1$, $k_2$, $k_3$, $V_b$, and $K_i$.}
    \label{fig:k myocardium}
\end{figure}

\begin{figure}[t]
    \centering
    \begin{subfigure}[t]{0.32\linewidth}
        \includegraphics[width=\linewidth,keepaspectratio]{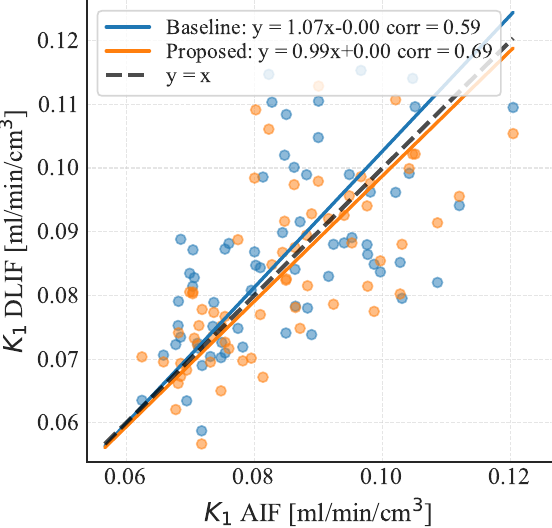}
        \caption{$K_1$}
        \label{fig:K1 brain}
    \end{subfigure}
    \begin{subfigure}[t]{0.32\linewidth}
        \includegraphics[width=\linewidth,keepaspectratio]{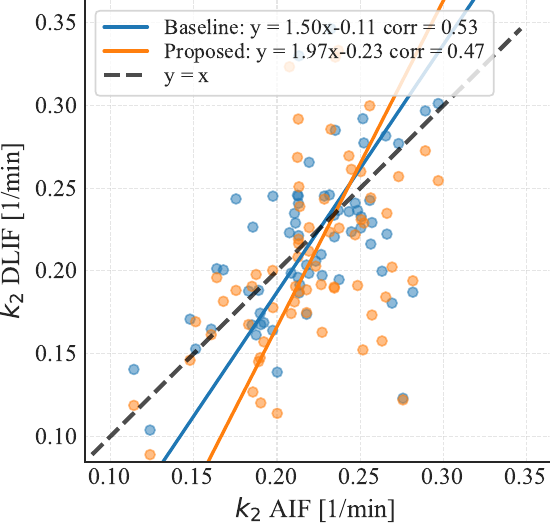}
        \caption{$k_2$}
        \label{fig:k2 brain}
    \end{subfigure}
    \begin{subfigure}[t]{0.32\linewidth}
        \includegraphics[width=\linewidth,keepaspectratio]{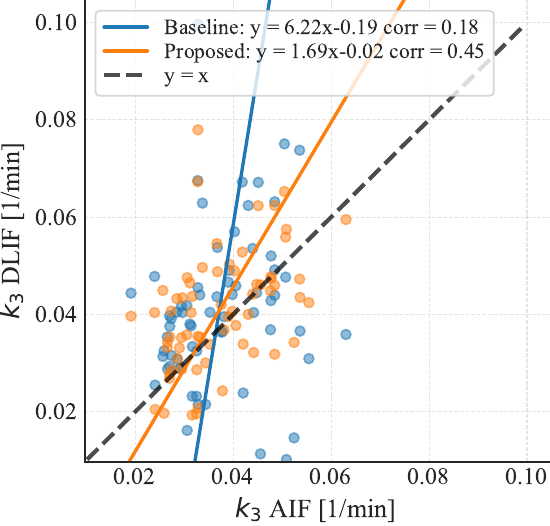}
        \caption{$k_3$}
        \label{fig:k3 brain}
    \end{subfigure}
    \begin{subfigure}[t]{0.33\linewidth}
        \includegraphics[width=\linewidth,keepaspectratio]{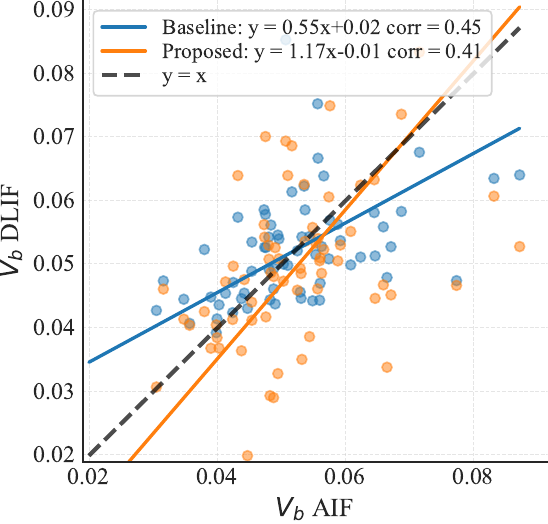}
        \caption{$V_b$}
        \label{fig:Vb brain}
    \end{subfigure}
    \begin{subfigure}[t]{0.33\linewidth}
        \includegraphics[width=\linewidth,keepaspectratio]{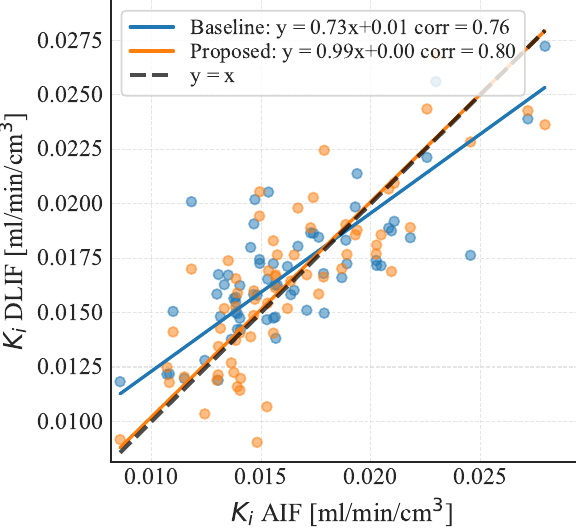}
        \caption{$K_i$}
        \label{fig:Ki brain}
    \end{subfigure}

    \caption[Estimated brain kinetic parameters using irreversible two-tissue
    compartment model.]%
    {Scatterplots summarizing kinetic parameters estimated using an irreversible
    two-tissue compartment model in the brain region. Points represent each region
    in each mouse sample, colored by the model producing the coefficient. From top left
    to bottom right: $K_1$, $k_2$, $k_3$, $V_b$, and $K_i$.}
    \label{fig:k brain}
\end{figure}

Further kinetic modeling results using the predicted input functions from both models are
summarized in Figures~\ref{fig:k myocardium} and~\ref{fig:k brain}. The kinetic
parameters were estimated using an irreversible two-tissue compartment
model~\cite{sokoloff1977kineticmodeling} in two different regions: myocardium and brain.
The estimated parameters were compared with those obtained using the measured AIF and
DLIF. The quantile--quantile plots show the distribution of the kinetic parameters
estimated using orthogonal regression.

Compared to the Patlak analysis presented in the main manuscript, the kinetic parameters
derived from the two-tissue compartment model are less stable because the full
measurement period is used in the fitting, and not only the linear phase. Outliers are
therefore removed if they lie more than three standard deviations away from the mean of
each parameter distribution. Once an outlier is detected, it is removed from all
parameters in the respective region for consistency. This results in \num{64} and
\num{66} samples remaining for the myocardium and brain regions, respectively.

The observed myocardium kinetic parameters (Figures~\ref{fig:k myocardium} and
\ref{fig:Ki myocardium}) follow the reference distribution well, with larger spread in
$k_2$, $k_3$, and $V_b$, as well as outliers not removed by the three standard deviation
rule in $k_2$ and the blood volume fraction. Influx rates
(Figure~\ref{fig:Ki myocardium}) were higher than previously
reported~\cite{kreissl2011myocardium,wong2011myocardium}. This is likely due to shorter
fasting times in our study, where longer fasting times are known to reduce influx rates
in the myocardium~\cite{kreissl2011myocardium,wong2011myocardium}.

The parameter distributions in the brain (Figure~\ref{fig:k brain}) had fewer outliers,
and similar to the myocardium, $k_2$, $k_3$, and $V_b$ had larger spread than the uptake
and influx rate. The ranges of the brain kinetic parameters are in line with previous
literature~\cite{alf2013fdg}.

\end{document}